\documentclass[bibyear]{aa} 

\usepackage{color}

\usepackage{graphicx}
\usepackage{txfonts}
\begin{document}

   \title{Gaia-DR2 extended kinematical maps}
   \subtitle{Part I: Method and application}

   \author{M. L\'opez-Corredoira\inst{1,2}, F. Sylos Labini\inst{3,4,5}}

   \institute{$^1$ Instituto de Astrof\'\i sica de Canarias, 
E-38205 La Laguna, Tenerife, Spain\\
$^2$ Departamento de Astrof\'\i sica, Universidad de La Laguna,
E-38206 La Laguna, Tenerife, Spain\\
 $^3$ Museo Storico della Fisica
          e Centro Studi e Ricerche Enrico Fermi, 
          Compendio del Viminale, 00184 Rome, Italy\\
$^4$ Istituto dei Sistemi Complessi, Consiglio Nazionale
          delle Ricerche,  00185 Roma, Italia \\ 
$^5$ Istituto Nazionale Fisica Nucleare, Unit\`a Roma 1, Dipartimento di
          Fisica, Universit\'a di Roma ``Sapienza'', 00185 Roma, Italia}

   \date{Received xxxx; accepted xxxx}

  
  \abstract
  {
  The Gaia Collaboration has used Gaia-DR2 sources with six-dimensional (6D) phase space information to derive kinematical maps within 5 kpc of the Sun, which is a reachable range for stars with relative error in distance lower than 20\%.
  }
   {Here we aim to extend the range of distances by a factor of two to three, thus adding  
the range of Galactocentric distances between 13 kpc and 20 kpc to the previous maps, with their corresponding error and root mean square values.}
{We make use of the whole sample of stars of Gaia-DR2 including radial velocity measurements, which consists in  
more than seven million sources, and we apply a 
statistical deconvolution 
of the parallax errors based on the Lucy's inversion method  of the
 Fredholm integral equations of the first kind, without assuming any prior.
}
{The new extended maps provide lots of new  and corroborated 
information about the disk kinematics: significant 
departures of circularity in the mean orbits with radial Galactocentric velocities between
-20 and +20 km/s and vertical velocities between -10 and +10 km/s; 
variations of the azimuthal velocity with position; asymmetries between the 
northern and the southern Galactic hemispheres, especially towards the 
anticenter that includes a larger azimuthal velocity in the south; and others.}
{These extended kinematical maps can be used to investigate the 
different dynamical models of our Galaxy, and we will present our own analyses in the forthcoming second part of this paper. At present, it is evident that
the Milky Way is far from a simple  stationary configuration in rotational equilibrium, but is characterized by streaming motions in all velocity components with
conspicuous velocity gradients.}
  

\titlerunning{Gaia-DR2 extended kinematics}
\authorrunning{L\'opez-Corredoira \& Sylos Labini}

   \maketitle
%

\section{Introduction}

The morphology of the stellar disk of our Galaxy has been explored in
recent years (e.g., Juric et al. 2008; Reyl\'e et al. 2009; Mateu et al. 2011; 
Polido et al. 2013; L\'opez-Corredoira \& Molg\'o 2014; Xu et al. 2015; 
Bovy et al. 2016), which has led to a 
global picture of the density laws with their
exponential radial and vertical dependences, 
together with their corresponding scale length and scale height, asymmetries, flare, warp, subdivision of a thin+thick component, 
and other features. 
The chemical abundances of the
thin and thick disks were also widely explored (e.g., Rong et al. 2001; Ak et al. 2007;
Casagrande et al. 2011; Haywood et al. 2013; Hayden et al. 2015) showing the different gradients of metallicities and alpha-enhancement.
However,  the
kinematics of the disk is not so well known, and we have only very recently begun
to gather relevant information away from the solar neighbourhood; see 
for example, Bond et al. (2010),
Siebert et al. (2011); Williams et al. (2013); Carrillo et al. (2018); 
Wang et al. (2018) and the set of three papers by one of the current authors dedicated to the azimuthal, vertical, and radial Galactocentric components, respectively: L\'opez-Corredoira (2014); L\'opez-Corredoira et al. (2014); L\'opez-Corredoira \& Gonz\'alez-Fern\'andez (2016).

 The Gaia
mission of the European Space Agency (Gaia Collaboration 2016) is leading us into a new era of the study of the kinematics of our Galaxy. Gaia offers
the position of each star in six-dimensional (6D) phase space: three dimensions of spatial information plus three dimensions of velocity.
Gaia data provide accurate distance determination for nearby sources, but
the errors increase with distance from us. 
For this reason, the Gaia Collaboration
(2018b; hereafter G18) and Kawata et al. (2018) carried out a kinematical analysis of the  second data release (DR2) of
the Gaia sample including radial velocities only for Galactocentric radii $R\lesssim 13$ kpc, which is a reachable range for stars with relative error in distance lower than 20\%. 
However, there  is a large number of 
Gaia-DR2 stars beyond $R$=13 kpc that may allow for 
larger distances to be explored, where several interesting features, such 
as the warp, flare, and others,  are known to develop.  
Poggio et al. (2018) extended the 
analyzed region up to $R\lesssim 15$ kpc in combination with 2MASS
photometry, but only paid attention to the vertical motions. We can extend the
Gaia maps much more than that limit: indeed 
our aim here is to provide maps of velocities 
  up to $R\approx 20$ kpc, using a technique to recover information on the 3D motions through a deconvolution of the Gaussian errors in the parallaxes, applicable even for large errors.
In a forthcoming work, we will consider the dynamical models that may explain 
the kinematical properties discussed here.

The  paper is organized  
as follows: in Sect. \ref{.data}, we describe the 
Gaia data  that 
we use in this paper; in Sect.  \ref{.galvel}, we explain the method for the conversion of heliocentric
to Galactocentric coordinates and for the deconvolution used to obtain information even
with large errors; the results of the application of the method to the real data is described
in Sect.  \ref{.results};  a discussion to sum up the paper,  with our 
conclusions, is  given in the
final section.


\section{Gaia-DR2 data}
\label{.data}

We use here the data of the second Gaia data release (Gaia DR2; Gaia Collaboration 2018a) for the stars with available radial heliocentric velocities: 7\,224\,631 sources, from which we take only those with
parallax error less than 100\% (7\,103\,123 sources), 
observed with a Radial Velocity Spectrometer (RVS; Cropper et al. 2018) 
that collects medium-resolution spectra
(spectral resolution $\frac{\lambda }{\Delta \lambda}\approx 11\,700$)
over the wavelength range 845-872 nm centered on the calcium triplet region.
This radial velocity data set contains the median radial velocities,
averaged over the 22-month time span of the observations. The sources are nominally brighter than twelfth magnitude in the G$_{RVS}$ photometric band, most having magnitudes brighter than 13 in 
the $G$ filter.

Radial velocities are only reported for stars with effective
temperatures in the range 3550-6900 K. The uncertainties of the radial velocities are: 0.3 km/s at G$_{RVS}<8$, 0.6 km/s at G$_{RVS}$=10, and
1.8 km/s at G$_{RVS}$= 11.75; plus systematic radial velocity errors 
of $< 0.1$ km/s at G$_{RVS}< 9 $ and  0.5 km/s at G$_{RVS}$= 11.75.
For details on 
radial velocity data processing and the properties and validation
of the resulting radial velocity catalogue, see Sartoretti
et al. (2018) and Katz et al. (2018). The set of standard stars that
was used to define the zero-point of the RVS radial velocities is
described in Soubiran et al. (2018).

Here we do not consider any possible zero-point bias in the parallaxes
of Gaia-DR2 as found by some authors (Lindegren et al. 2018; Arenou et al. 2018; Stassun \& Torres 2018; Zinn et al. 2018), except
in Sect.  \ref{.0point} where we repeat some of the key calculations 
with a non-zero value of the zero-point in order to show that its 
effect on our main results is negligible.


\section{Obtaining the Galactocentric velocity}
\label{.galvel}

The information provided by the Gaia-DR2 catalog for each star contains the parallax $\pi $, the Galactic coordinates ($\ell$, $b$), the radial velocity $v_r$ and the two proper motions in equatorial coordinates $\mu _\alpha \cos \delta $ and $\mu _\delta $ (which can be transformed
into Galactic coordinates as $\mu _\ell\cos b$ and $\mu _b$). These six variables allow for each star to be placed in the 6D phase space: 3D spatial + 3D velocity.

The Galactocentric position in cylindrical coordinates has the following three components: 
the Galactocentric distance $R$, the Galactocentric azimuth $\phi $ \footnote{We define it such that 
$\phi _\odot =0$, $X_\odot =R_\odot $; we note that the definition used by
G18 is different as they have fixed $\phi _\odot =0$, $X_\odot =-R_\odot $.}, and the vertical distance $Z$. 
Analogously, the Galactocentric velocity in cylindrical coordinates has the following three components: the 
radial velocity $V_R$, the azimuthal velocity $V_\phi $, and the 
vertical velocity $V_Z$. The convention for the sign for the 
Galactocentric radial velocity is positive when directed outwards.
The conversion formulas are the following 
(L\'opez-Corredoira \& Gonz\'alez-Fern\'andez 2016):
\begin{equation}
\label{transpos}
r=\frac{1\ {\rm AU}}{\pi }
\end{equation}\[
X(r,\ell, b)=R_\odot -r\cos \ell \cos b 
,\]\[
Y(r,\ell, b)=r\sin \ell \cos b
,\]\[
Z(r, b)=Z_\odot + r\sin b
,\]\[
R(r,\ell, b)=\sqrt{X^2+Y^2}
,\]\[
\phi (r,\ell, b)=\tan ^{-1}\left(\frac{Y}{X}\right) \ \ \left({\rm between}\ \frac{\pi}{2}\ {\rm and}\ \frac{3\pi}{2}\ {\rm for}\ X<0\right) 
\;,\]
and
\begin{equation}
\label{transvel}
\vec{V}(r,\ell, b, v_r,\mu _\ell, \mu _b )=
\end{equation}\[
\left(\begin{array}{c}
V_R \\
V_\phi \\
V_z
\end{array}\right)=
\left(\begin{array}{ccc}
-\cos \phi  & \sin \phi  & 0 \\
\sin \phi  & \cos \phi  & 0 \\
0 & 0 & 1 
\end{array}\right)
\left(\begin{array}{c}
U_*+U_\odot \\
V_*+V_{g,\odot} \\
W_*+W_\odot
\end{array}\right)
,\]\[
\left(\begin{array}{c}
U_* \\
V_* \\
W_*
\end{array}\right)=
\left(\begin{array}{ccc}
\cos \ell \cos b  & -\sin \ell  & -\cos \ell \sin b  \\
\sin \ell \cos b   & \cos \ell  & -\sin \ell \sin b \\
\sin b & 0 & \cos b 
\end{array}\right)
\left(\begin{array}{c}
v_r \\
v_\ell \\
v_b
\end{array}\right)
,\]\[
v_\ell=r\,\mu _\ell \cos b      
,\]\[
v_b=r\,\mu _b
.\]
Here we adopt the same Galactic parameters used by the previous kinematic Gaia-DR2 analysis (G18): $R_\odot =8.34$ kpc (Reid et al. 2014), $Z_\odot =0.027$ kpc (Chen et al. 2001); $(U_\odot , V_\odot , W_\odot )=(11.1,12.24,7.25)$ km/s (Sch\"onrich et al. 2010); 
$V_{g,\odot}=V_{c,\odot}+V_\odot$, with the rotation speed $V_{c,\odot}=240$ km/s (Reid et al.
2014).


\subsection{Deconvolution of parallax errors}
\label{.deconv}

The main problem to solve is the deconvolution of the Gaussian errors with large root mean square (r.m.s) values.
The large dispersion of parallaxes (and hence the 
 large dispersion of distances) affect
 the value of mean distance at large heliocentric distances, which is 
indeed overestimated in a direct measurement with respect to the real one.  
 This occurs because the parallax errors grow with distance and  
because of a  selection effect: 
  most of the stars in  a  bin 
  centered at $r$ are indeed at a much lower distance 
  for the observational asymmetry, meaning that
there are much more stars at distances $<r$ than at distances $>r$.
The symmetric parallax uncertainties ($\sigma_ \pi $) 
produce asymmetric distance uncertainties, with a probability 
distribution function stretched toward larger distances: the 
largest the relative uncertainty on the parallax, the largest the asymmetry.

Let us explain this point in more detail.
The number of stars per unit parallax that we see in a given line of sight is $\overline{N}(\pi )$. However, this is not the true number of stars per unit parallax, $N(\pi )$, but its convolution with a Gaussian function that takes into account the measurement error, that is,
\begin{equation}
\overline{N}(\pi )=\int_0^\infty d\pi'\,N(\pi ')G_{\pi '}(\pi -\pi ')
\label{convol}
,\end{equation}
and 
\begin{equation}
G_\pi (x)=\frac{1}{\sqrt{2\pi }\sigma _\pi}e^{-\frac{x^2}{2\sigma _\pi ^2}}
.\end{equation}
We take as $\sigma_ \pi$ the average error on $\pi $ given by 
the Gaia-DR2 survey 
computed for the stars of each bin.
We note that $N(\pi )$ contains the dependence of the stellar density along the line of sight
multiplied by the selection factors: 
\begin{equation}
\label{Nselec}
N(\pi)=\omega \frac{\rho (1/\pi )}{\pi ^4}
\int _{-\infty}^{m_{G,lim}-5\log _{10}(1/\pi)-10-A_G (1/\pi)}dM_G\,\phi _G(M_G)
,\end{equation}
where $\omega $ is the covered angular surface of the line of sight (in stereoradians), 
$\rho (r)$ is the stellar density along the line of sight, $\phi _G(M_G)$ 
is the luminosity function in the $G$ filter as a function of the 
absolute magnitude $M_G$, 
$m_{G,lim}$ is the limiting maximum apparent magnitude of the
survey (in our case $m_{G,lim}$ is between 12 and 13) and $A_G(r)$ is the cumulative extinction out to a distance of $r$.

Equation (\ref{convol}) is a Fredholm integral equation of the first kind
  with kernel equal to $G_\pi (\pi -\pi ')$. The inversion of this equation can be carried out and
thus we can obtain a solution $N(\pi )$ from the observed data $\overline{N}(\pi )$.
For that, we use here an iterative Bayesian method called Lucy's method 
that is detailed in Appendix \ref{.Lucy}. 
The integrals are discretized with bins of width $\Delta \pi=0.01$ mas; the results do not change with the variation of $\Delta \pi$, but the resolution and the range of the explored range depend on it.
There are also other methods of Gaussian deconvolution in the literature appropriate for different kinds of physical problems (e.g., Masry \& Rice 1992; Fang et al. 1994; Ulmer 2010, 2013), but an iterative method is more appropriate here because it is more robust against noise.

Once we have obtained the deconvolved distribution of sources along the line of sight,
we estimate the mean heliocentric distance of all the stars included 
in a bin of parallax $\pi $ and the corresponding variance:
\begin{equation}
\label{rdeconv}
\overline{r}(\pi )=\frac{1}{\overline{N}(\pi )}
\int_0^\infty d\pi'\,\frac{N(\pi ')}{\pi '}G_{\pi '}(\pi -\pi ')
,\end{equation}
\begin{equation}
\label{variance}
\sigma _{\overline{r}}^2(\pi )=\left[\frac{1}{\overline{N}(\pi )}
\int_0^\infty d\pi'\,\frac{N(\pi ')}{\pi '^2}G_{\pi '}
(\pi -\pi ')\right]-\overline{r}(\pi )^2
.\end{equation}
For obtaining the variance in the positive and in the negative parts separately,
the expressions would be similar to Eqs. (\ref{convol}) and (\ref{variance})
but with integral limits (0,$1/\overline{r}(\pi )$] and ($1/\overline{r}(\pi )$,$\infty $),
respectively.

\begin{figure}
\vspace{1.5cm}
\centering
\includegraphics[width=8cm]{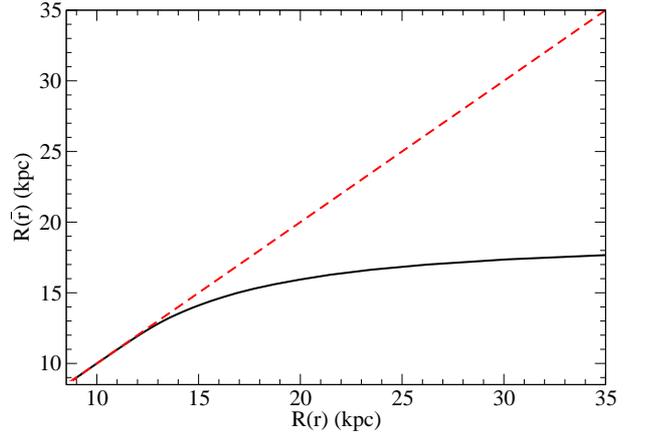}
\caption{The Galactocentric distance 
$R$, obtained after the deconvolution of the heliocentric distances 
through Eq. (\ref{rdeconv}), vs. $R$,  obtained from the  
heliocentric distances as $1/\pi $ 
(see Eq. (\ref{transpos})). 
For comparison,
the dashed line ($y=x$) shows how the Galactocentric distance $R$ 
should  behave without the deconvolution correction.
This result is obtained for the 
Gaia-DR2 data including $v_r$ measurements with the constraints:
$|\ell -180^\circ |<20^\circ$, $|b|<10^\circ $, $\frac{\pi}{\Delta \pi}>1$.}
\vspace{.2cm}
\label{Fig:Rdeconv}
\end{figure}

In Fig. \ref{Fig:Rdeconv} we  show how the Galactocentric distance $R$ changes 
when taking into account this deconvolution, for a particular line of
sight in the direction of the anticenter.
In the anticenter direction the correction is negligible for $R\lesssim 13$ kpc, but becomes very important beyond that radius. For the stars with measured 
$R\approx 35$ kpc ($r\approx 27$ kpc,
$\pi \approx \Delta \pi =0.037$ mas; that is, the case with 100\% error in heliocentric distance), we get that the mean average value of $R$ is indeed $\approx 18$ kpc. This gives us an idea of the limits of Gaia DR2, either with or without radial velocities:  we cannot go beyond $\approx 20$ kpc, and most of the stars with measured larger distance are indeed stars with much lower distance but with a large positive error.

It is worth stressing that the Lucy's
 method is model independent, that is, we can recover information 
 on the stellar distribution without introducing any prior. This is  necessary 
if one does not know
a priori that information. Other
 Bayesian methods
  used to deconvolve Gaussian errors
of Gaia parallaxes (e.g., Astraatmadja \& Bailer-Jones 2016; Bailer-Jones et al. 2018) 
make use of a prior on the density of the Milky Way to improve the determination of the distances, meaning that they are model dependent.

\begin{figure}
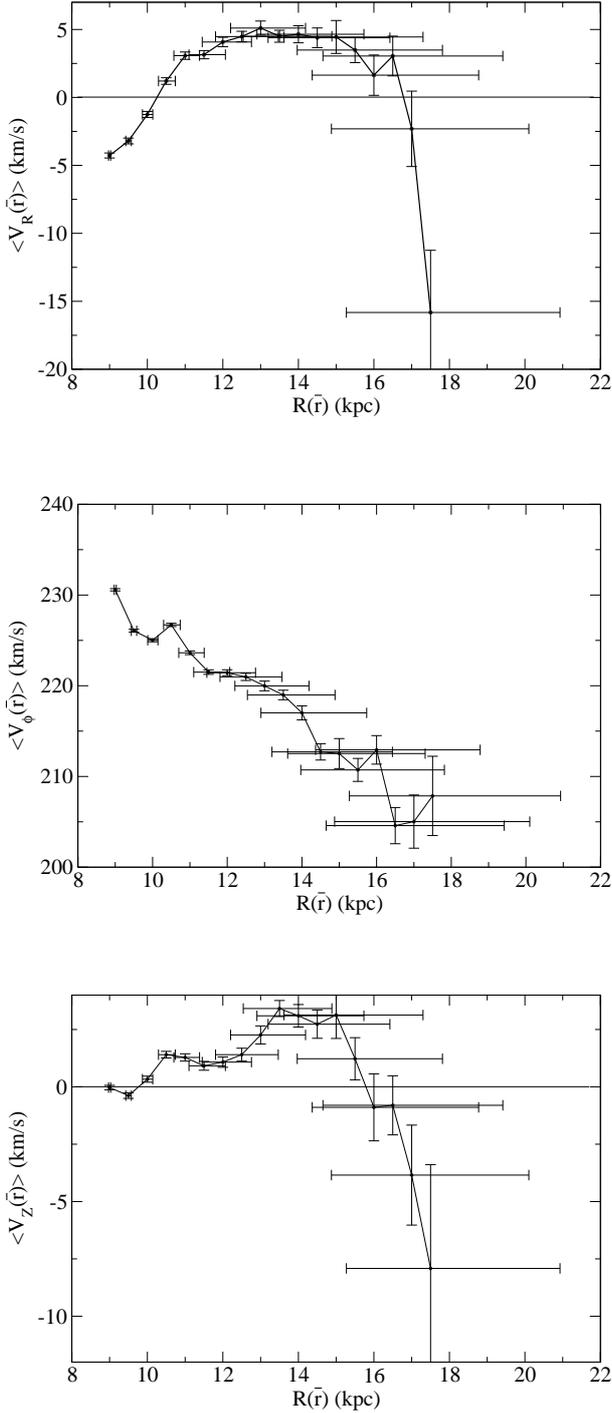

\vspace{1.5cm}
\centering
\includegraphics[width=8cm]{vr.eps}\\
\vspace{1cm}
\includegraphics[width=8cm]{vphi.eps}\\
\vspace{1cm}
\includegraphics[width=8cm]{vz.eps}
\caption{Median of Galactocentric radial (top panel)
 azimuthal (middel panel)  
 and vertical (bottom panel) 
 velocity as a function of $R(\overline{r})$ ($R$ 
 with a deconvolution correction according to Sect.  \ref{.deconv}).
This is the result 
for Gaia-DR2 data including $v_r$ measurements with the constraints:
$|\ell -180^\circ |<20^\circ$, $|b|<10^\circ $, $\frac{\pi}{\Delta \pi}>1$.
Vertical bars indicate the errors on the median velocity 
(without including the
uncertainty in distance). 
Horizontal bars indicate the
r.m.s. value of  the mean deconvolved $R$ (thus the indicated value 
is an average over that range of dispersion).}
\vspace{.2cm}
\label{Fig:vr}
\end{figure}

\begin{figure}
\vspace{1.5cm}
\centering
\includegraphics[width=8cm]{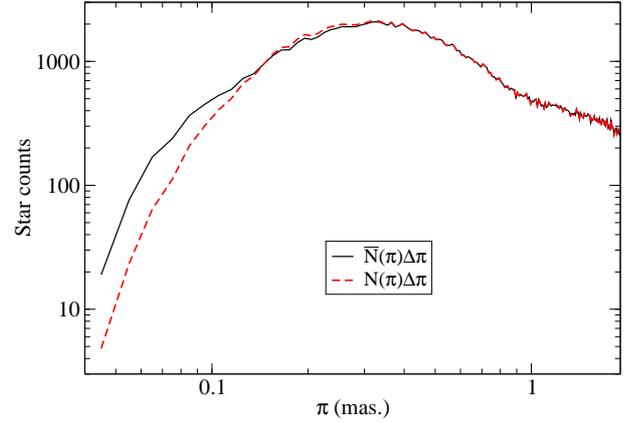}
\caption{
Log-log plot of the observed counts of stars $\overline{N}\Delta \pi$  
(with a bin width $\Delta \pi=0.01$ mas.)
as a function of the  parallax, together with the 
calculated star counts  $N\Delta \pi$ after  the
deconvolution was applied, for the
 Gaia-DR2 with measured radial velocity and 
 with the constraints:
$|\ell -180^\circ |<20^\circ$, $|b|<10^\circ $, $\frac{\pi}{\Delta \pi}>1$.}
\vspace{.2cm}
\label{Fig:npi}
\end{figure}

Unfortunately, we cannot apply the same method of inversion to obtain the different
component of $\vec{V}$:
Lucy's method only works with positive magnitudes, and other analytical inversion
methods are less robust  as 
there is a degeneracy of solutions in a combination of positive
and negative values that limits the inversion possibilities. Nonetheless,
we can calculate the  distribution of the velocity $\vec{V}$ 
as a function of $R$ 
with the deconvolution correction according to the above prescription: that is, 
$\vec{V}(\overline{r},\ell, b, v_r,\mu _\ell, \mu _b )$ as a function of
$R(\overline {r},\ell ,b$), using the transformations
of Eqs. (\ref{transpos}) and (\ref{transvel}); in each bin with mean distance $\overline {r}$, we recalculate
the velocities as a function of the recalculated Galactocentric distance.
An example is given in Fig. \ref{Fig:vr}, for the $R$, $\phi, $ and $Z$ 
components, respectively, in the direction towards the 
anticenter. We also show in Fig. \ref{Fig:npi}
the number of stars per unit parallax
before and after deconvolution; as can be observed, the application of
the method significantly decreases the number of stars with low $\pi $.
We note that the plotted velocities are the
median values within a range of distances $R$ whose r.m.s. value 
is indicated by 
the horizontal error bars. 

\subsection{Monte Carlo simulations to test the inversion technique}
\label{.monte}

In order to test the method explained in Sect.  \ref{.deconv}, we  
present here the results of several Monte Carlo simulations.
In our numerical experiments 
we assume a simple Galactic model with stellar density 
\begin{equation}
\rho (R,z)=\rho _0\times e^{-\frac{R-R_\odot }{2\ {\rm kpc}}}
e^{-\frac{|z|}{0.3\ {\rm kpc}}} \,,
\end{equation}
and with a luminosity function $\phi (M)$ for the disk  taken from
Bahcall \& Soneira (1980).
These are rough measurements of the density and of 
the luminosity function, but here they are not used with any purpose of analysis of the true Milky Way distribution,
but only to construct simple tests of the Lucy's method.
We take $m_{\rm max}=13$, 
$R_\odot =8.3$ kpc  as limiting magnitude, and we assume two different radial velocity distributions 
as a function of the heliocentric distance $r$: 
i) $V_R({\rm km/s})=13-R({\rm kpc})$;
ii) $V_R({\rm km/s})=10\,\sin(1.5\times 
[\sqrt{R/(1\ {\rm kpc})}-\sqrt{R_\odot/(1\ {\rm kpc})]})$.
We also take the azimuthal velocity $V_\phi({\rm km/s})=240$ for $R<10$ kpc,
$V_\phi({\rm km/s})=240\sqrt{(10\ {\rm kpc})/R}$ for $R\ge 10$ kpc,
and we take a vertical velocity $V_z({\rm km/s})=20\,\sin[R/(10\ {\rm kpc})]$. 
We apply a deconvolution of Gaussian errors assuming $\sigma _\pi =0.04$ milliarcseconds.
We assume an error of heliocentric velocity measurements of 2 km/s (i.e., 
of the order of magnitude of the value 
between 1.4 and 3.7 km/s for the faintest sources according to G18)
and an error on proper motions of 0.07 mas/yr, typical of Gaia-DR2 (G18).

\begin{figure}[htb]
\vspace{1.5cm}
\centering
\includegraphics[width=8cm]{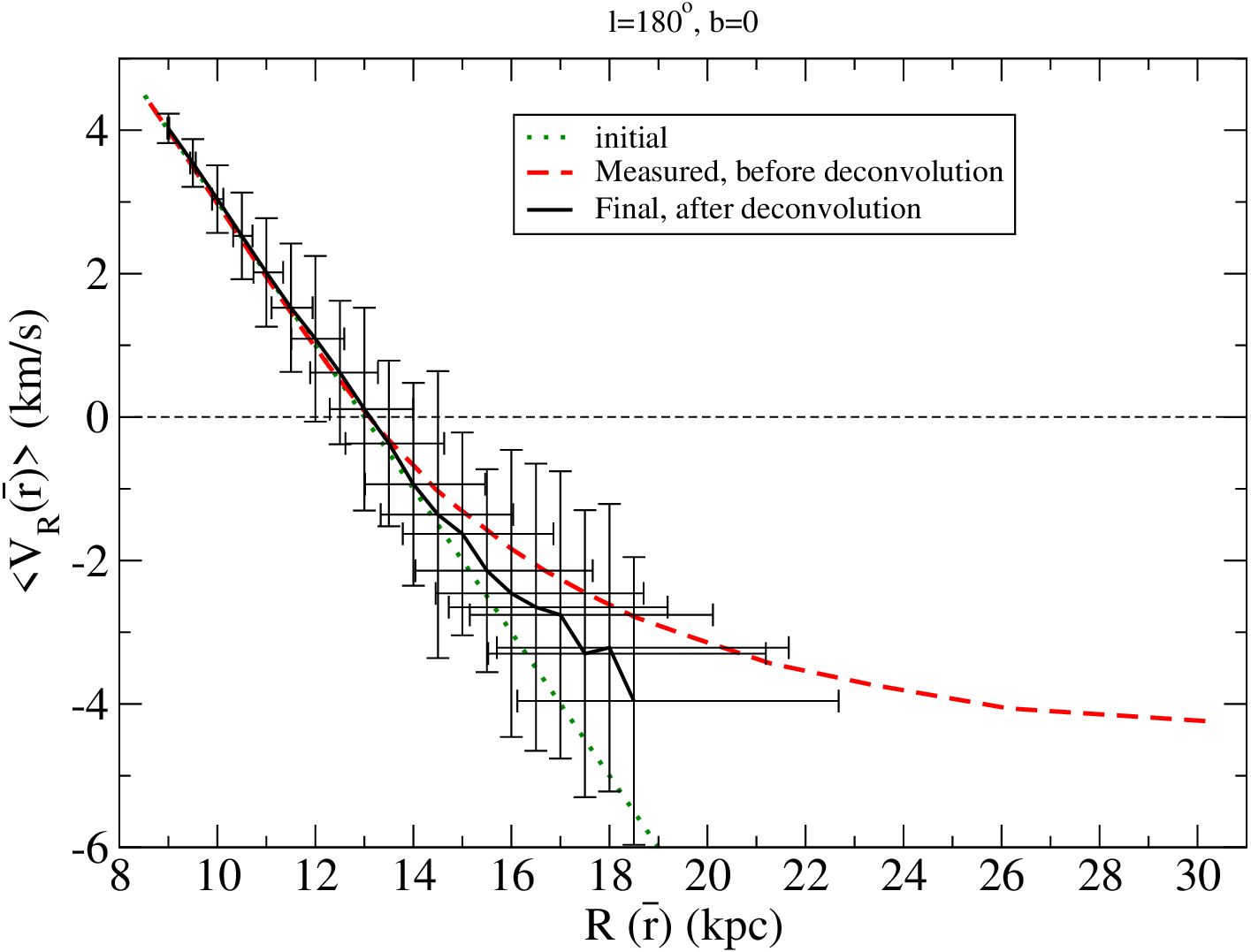}\\
\vspace{1cm}
\includegraphics[width=8cm]{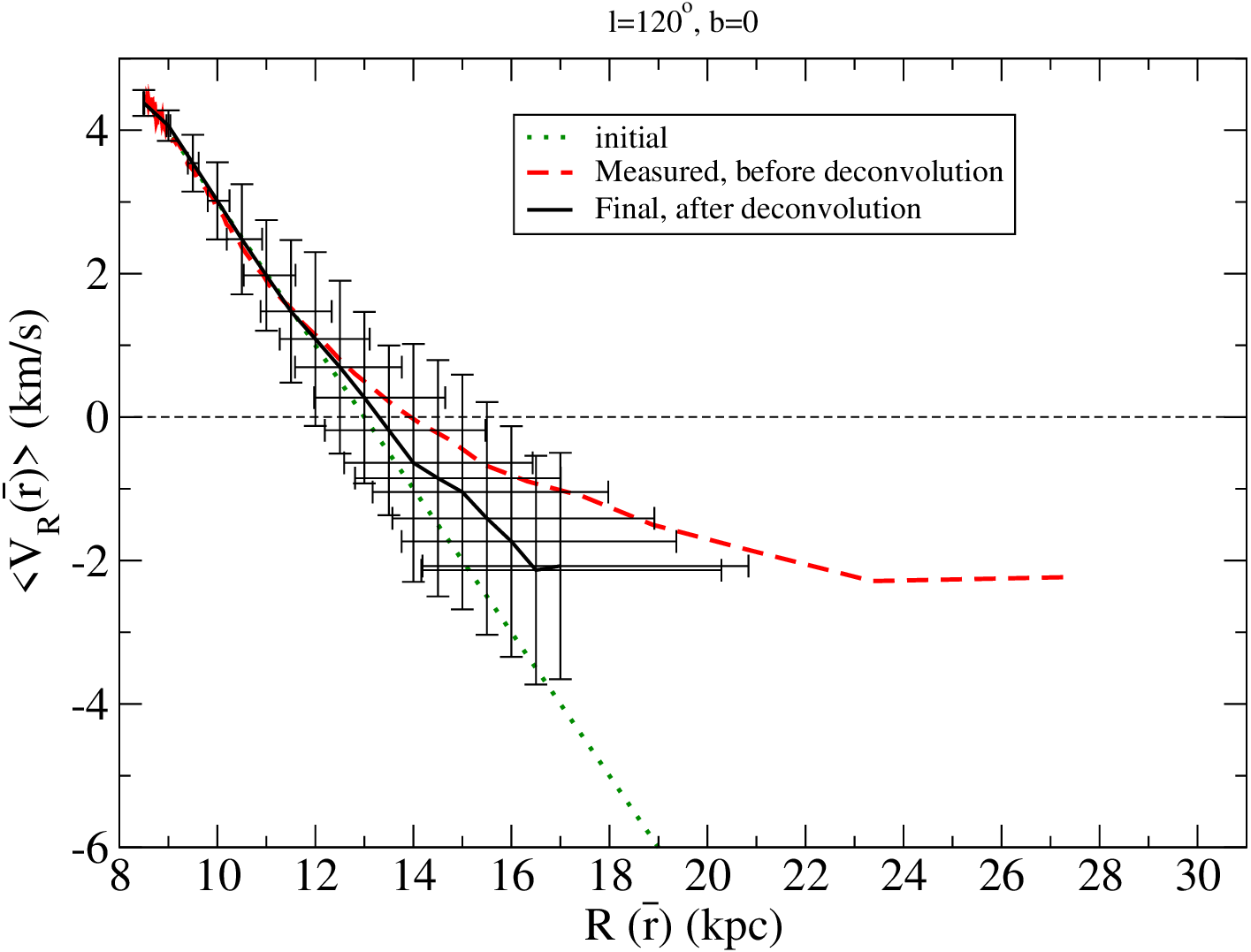}\\
\vspace{1cm}
\includegraphics[width=8cm]{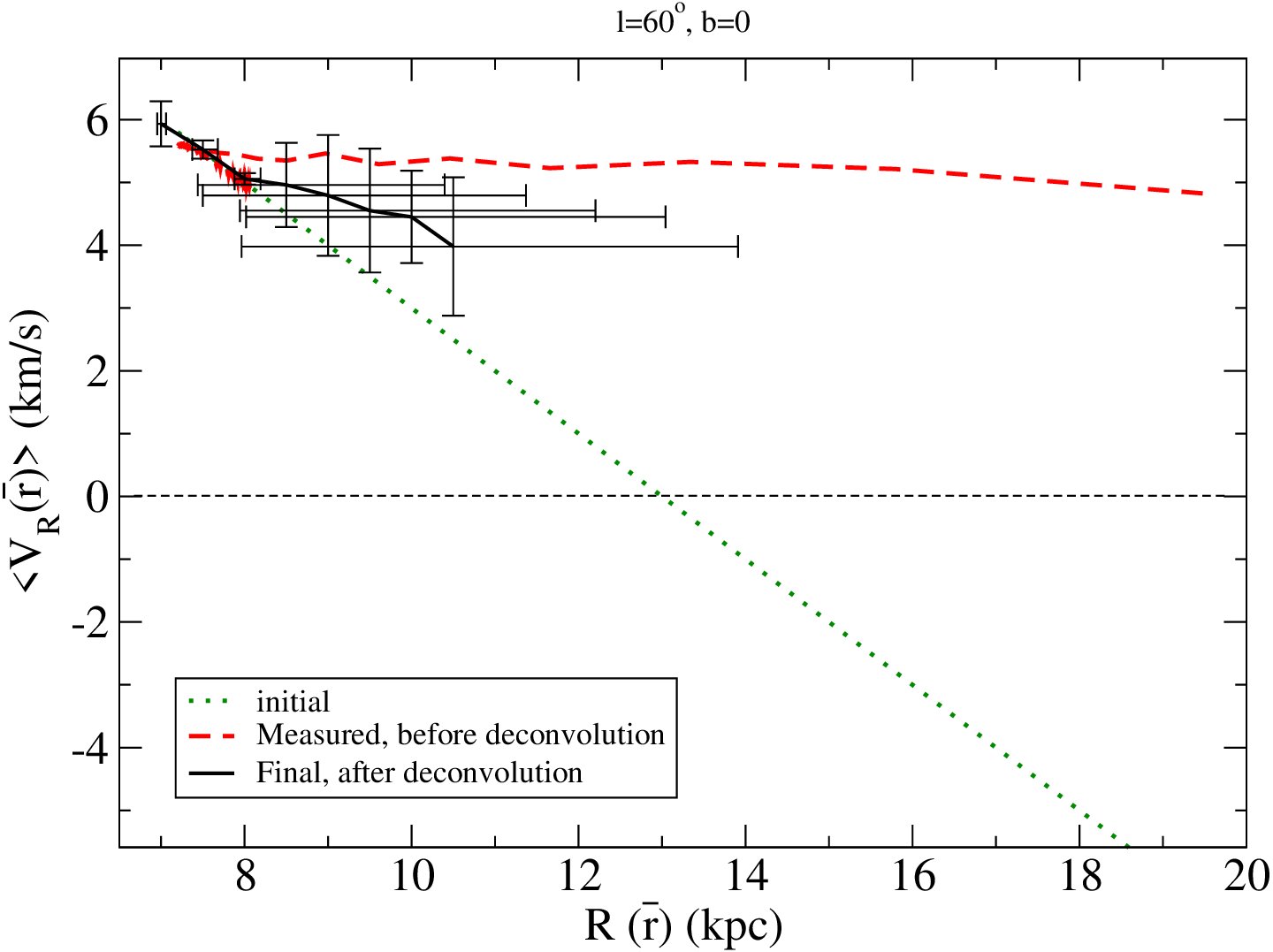}
\caption{Monte Carlo simulation of $V_R$: with an initial 
$V_R({\rm km/s})=13-R({\rm kpc})$, the direct measurement ``before deconvolution'', and the ``final'' result once the deconvolution has been applied. Horizontal error bars indicate the
r.m.s. value  of  the mean deconvolved $R$ (thus the indicated value is an 
average over that range of dispersion). The three panels are for the directions in the planes 
$\ell =180 ^\circ$, $\ell =120^\circ $ , and $\ell =60^\circ $,  
respectively.}
\vspace{.2cm}
\label{Fig:simul1}
\end{figure}

\begin{figure}[htb]
\vspace{1.5cm}
\centering
\includegraphics[width=8cm]{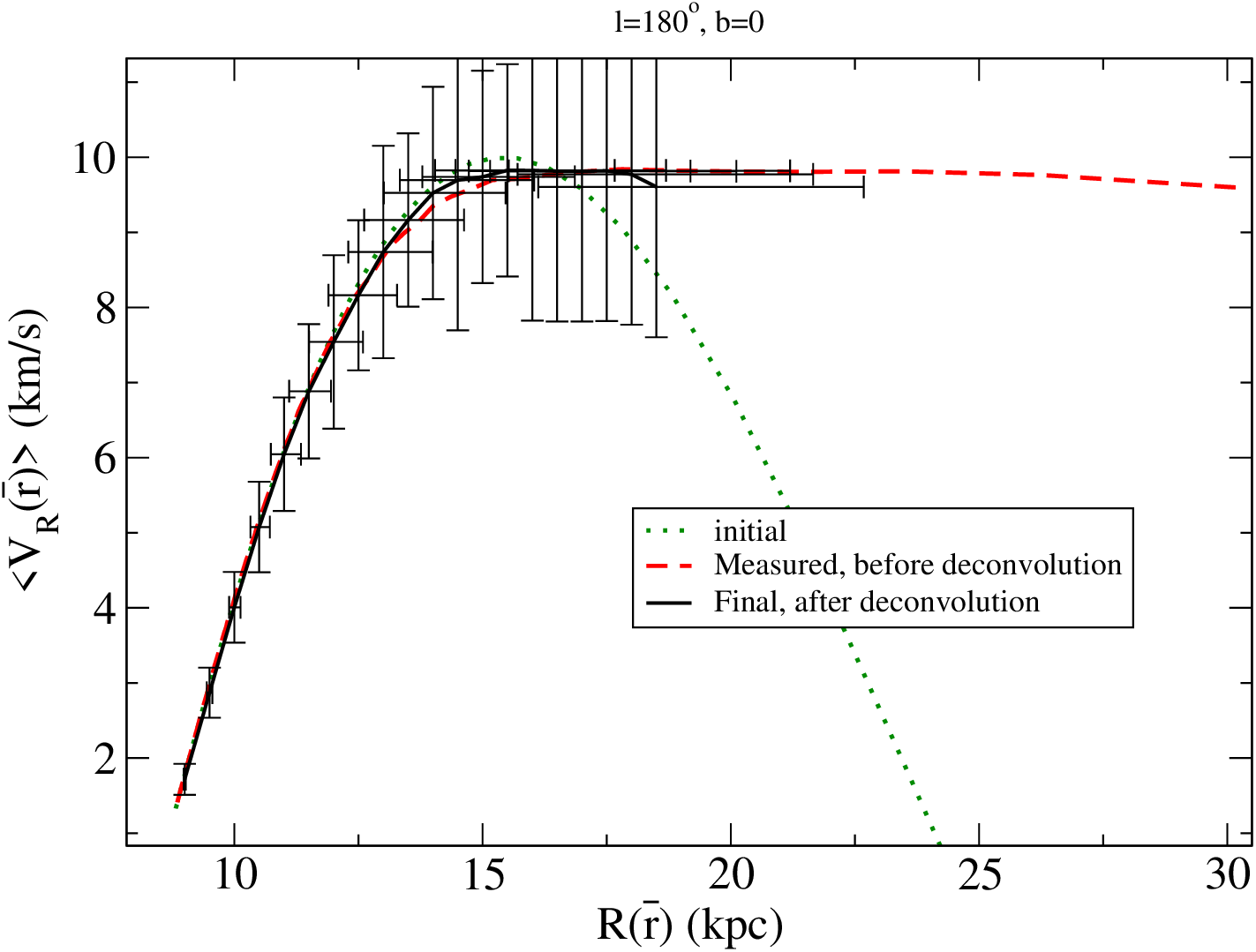}\\
\vspace{1cm}
\includegraphics[width=8cm]{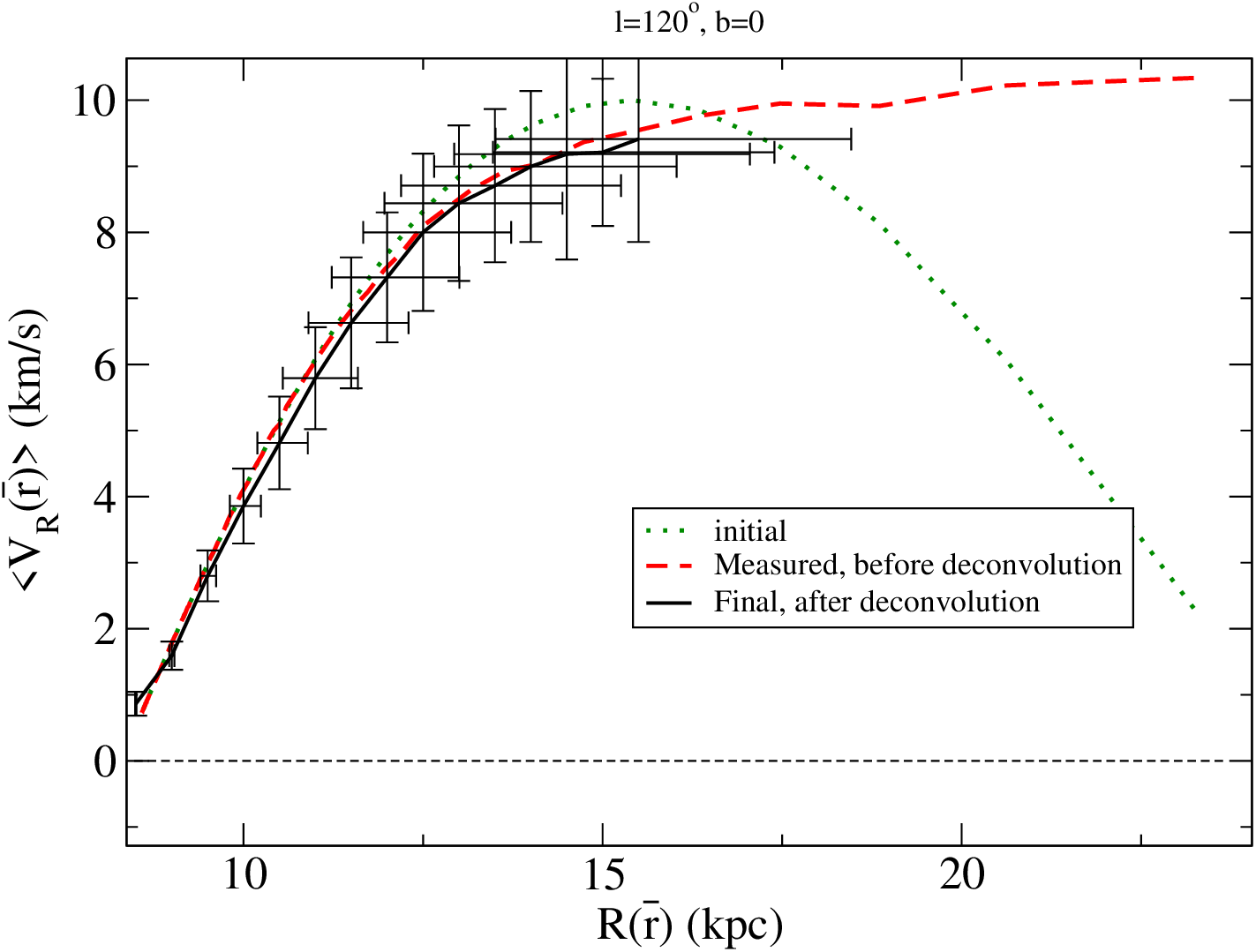}\\
\vspace{1cm}
\includegraphics[width=8cm]{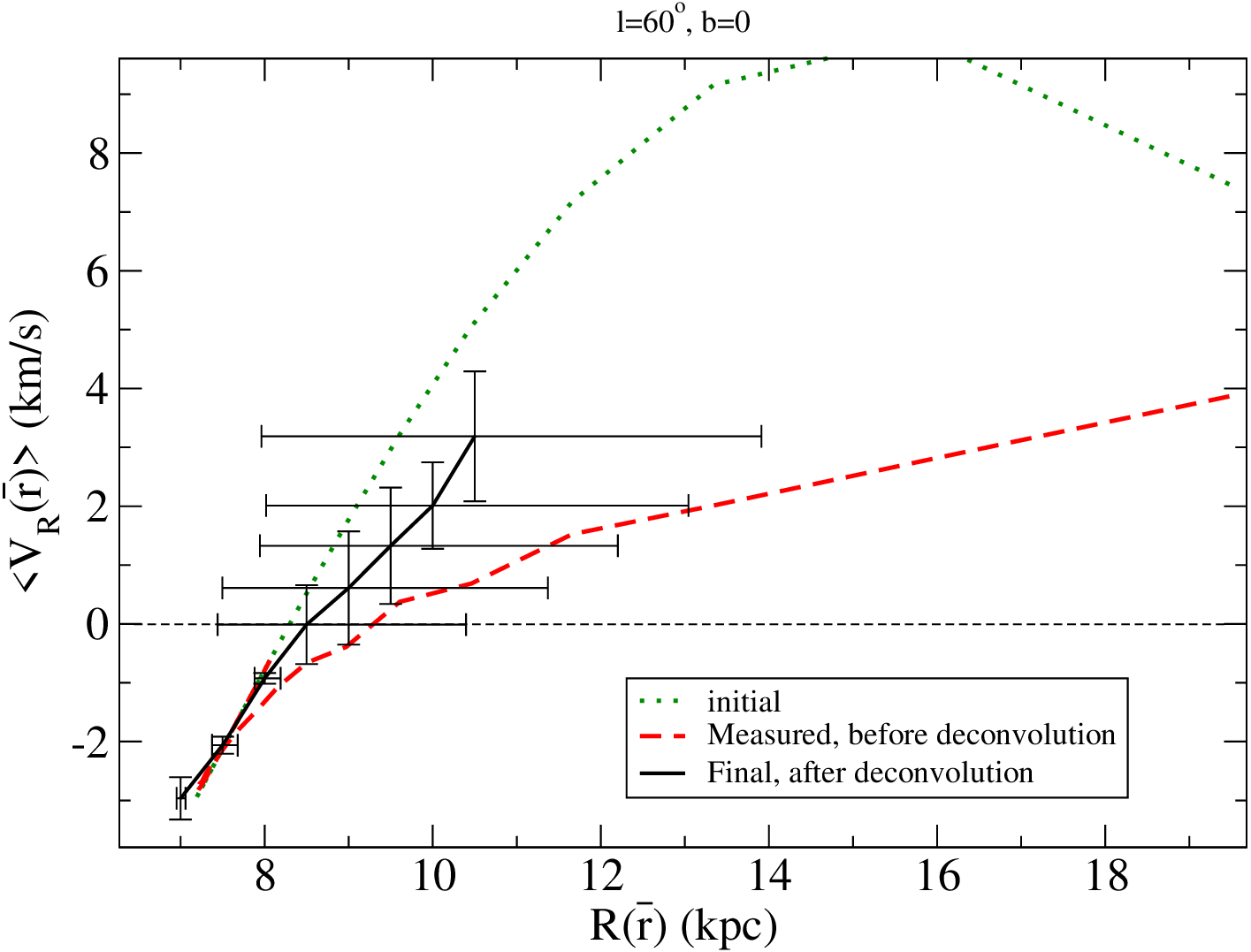}
\caption{As in Fig. \ref{Fig:simul1} but with initial velocity 
$V_R({\rm km/s})=10\,\sin(1.5\times 
[\sqrt{R/(1\ {\rm kpc})}-\sqrt{R_\odot/(1\ {\rm kpc})]})$.}
\vspace{.2cm}
\label{Fig:simul2}
\end{figure}

\begin{figure}[htb]
\vspace{1.5cm}
\centering
\includegraphics[width=8cm]{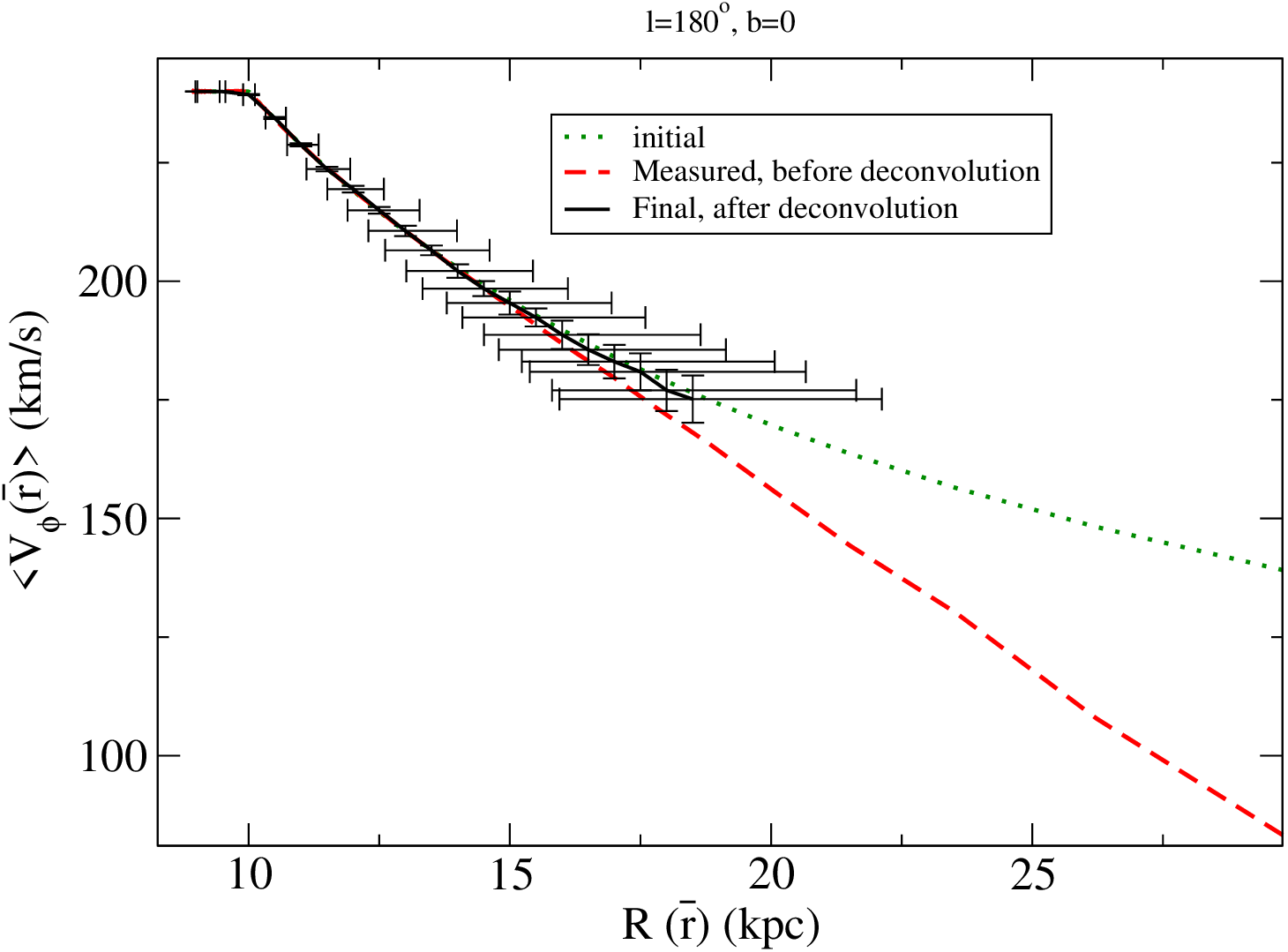}\\
\vspace{1cm}
\includegraphics[width=8cm]{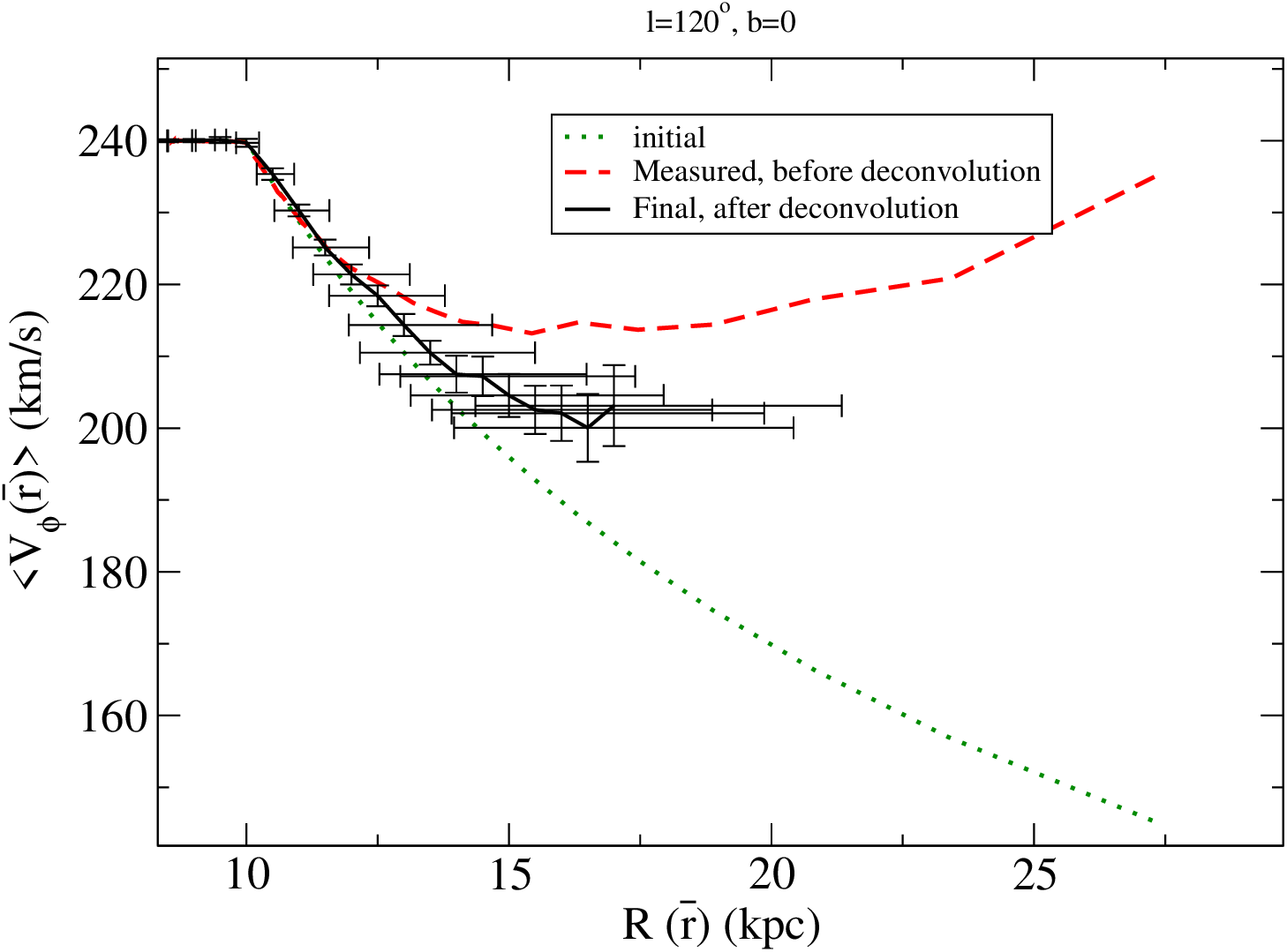}\\
\vspace{1cm}
\includegraphics[width=8cm]{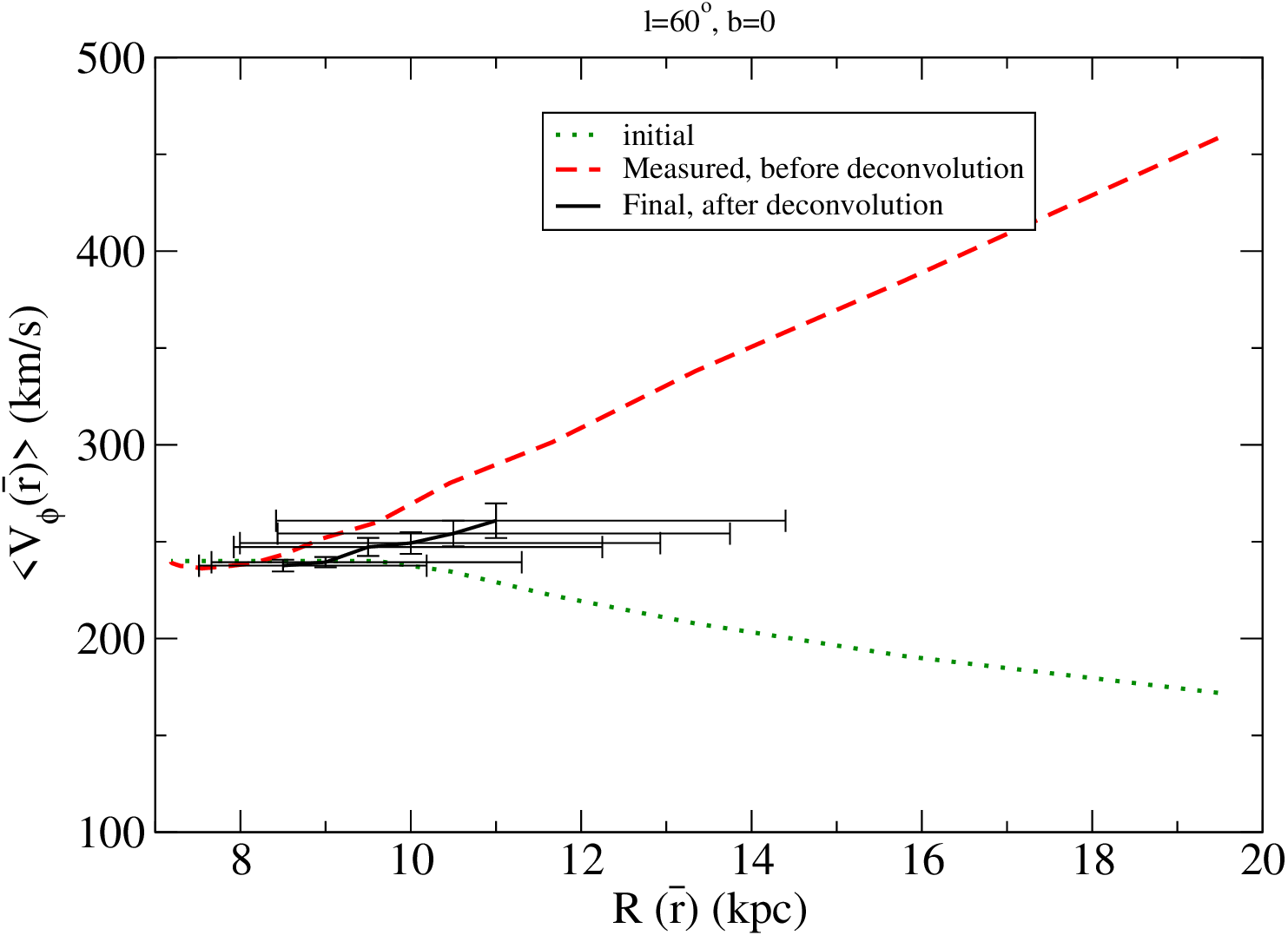}
\caption{Monte Carlo simulation of $V_\phi$. Horizontal error bars indicate the
r.m.s. value  of  the mean deconvolved $R$ (thus the indicated value is an average over that range of dispersion). The three panels are for the directions in the plane
$\ell =180 ^\circ$, $\ell =120^\circ $ and $\ell =60^\circ $, respectively.}
\vspace{.2cm}
\label{Fig:simul3}
\end{figure}

\begin{figure}[htb]
\vspace{1.5cm}
\centering
\includegraphics[width=8cm]{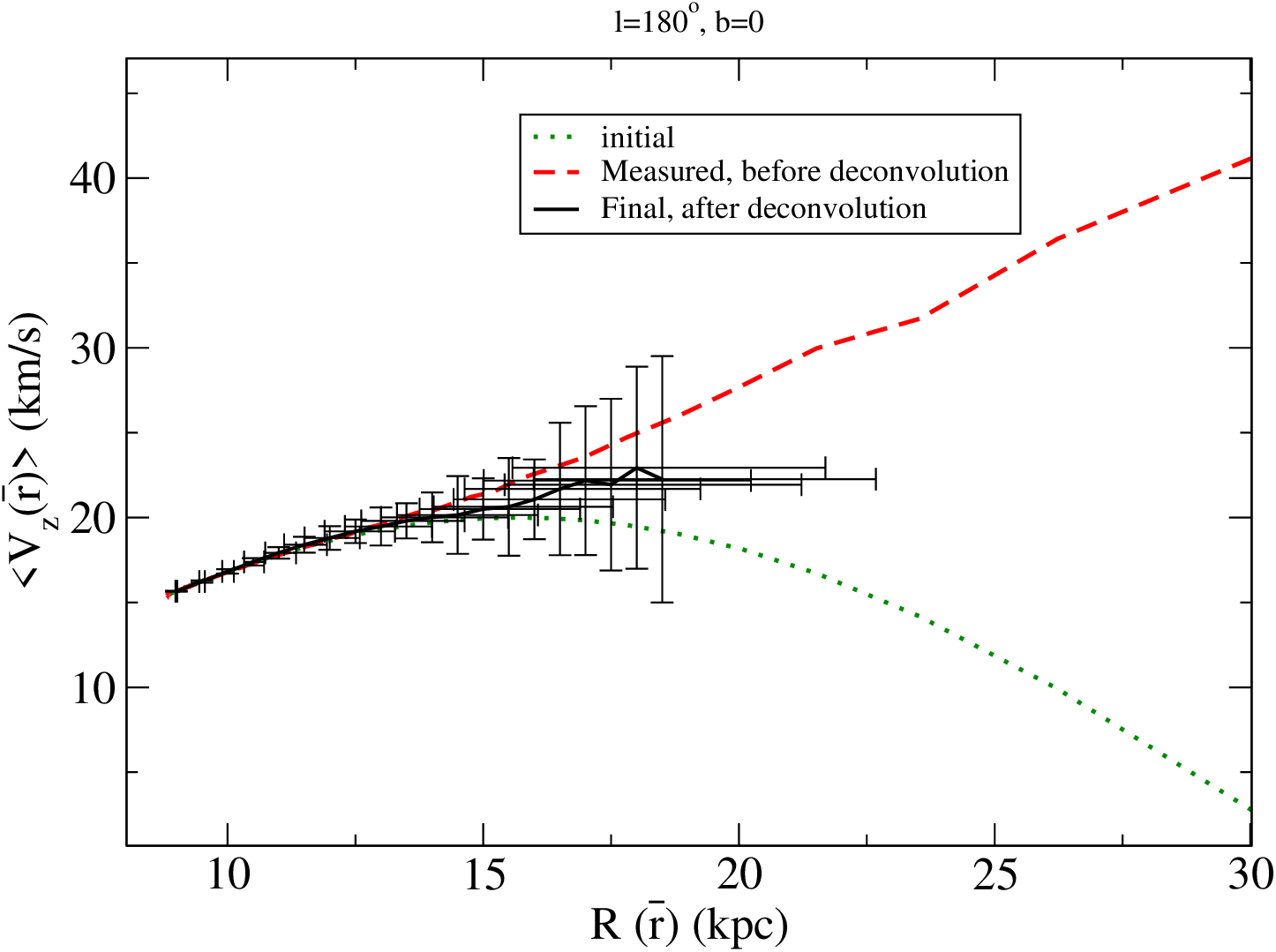}\\
\vspace{1cm}
\includegraphics[width=8cm]{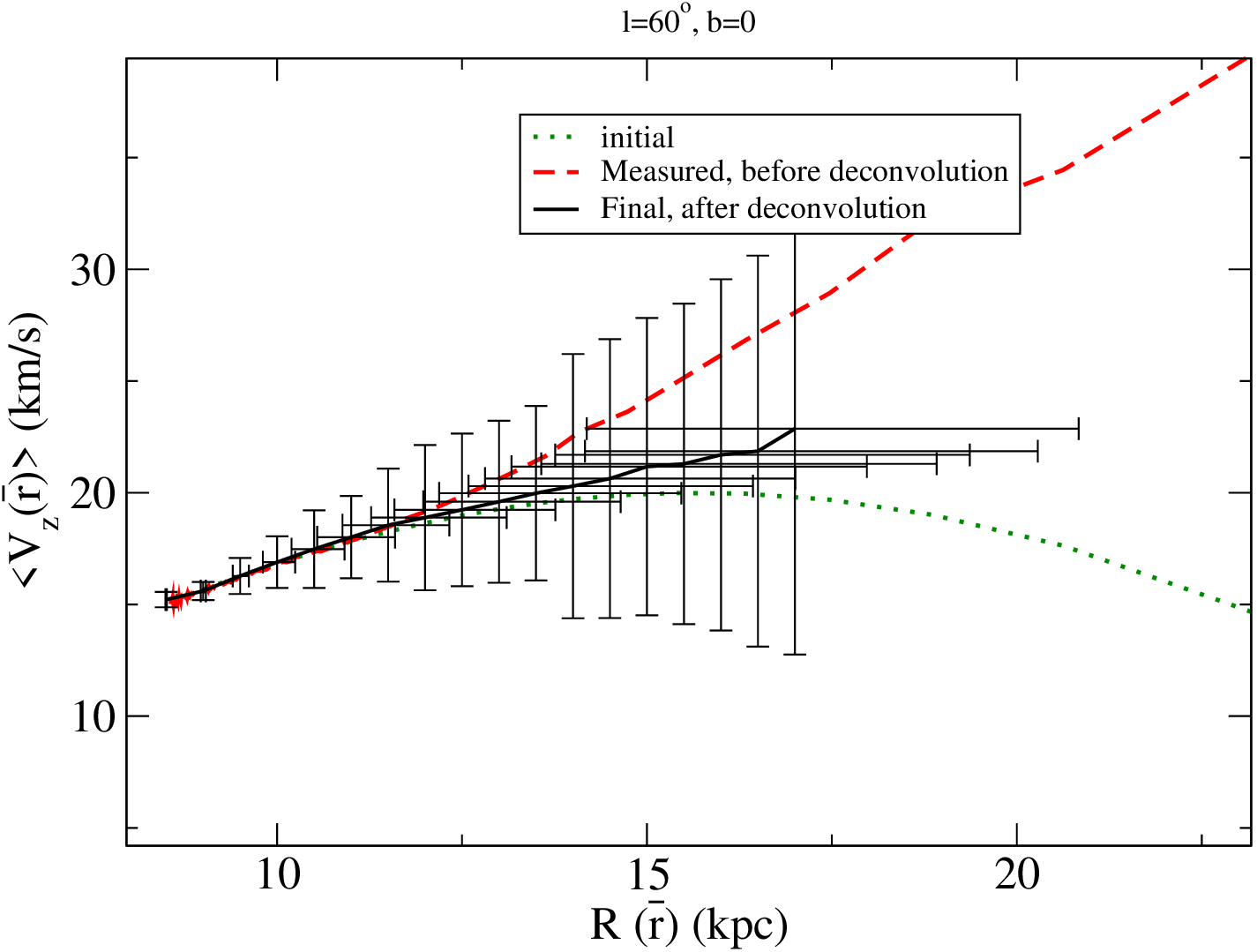}\\
\vspace{1cm}
\includegraphics[width=8cm]{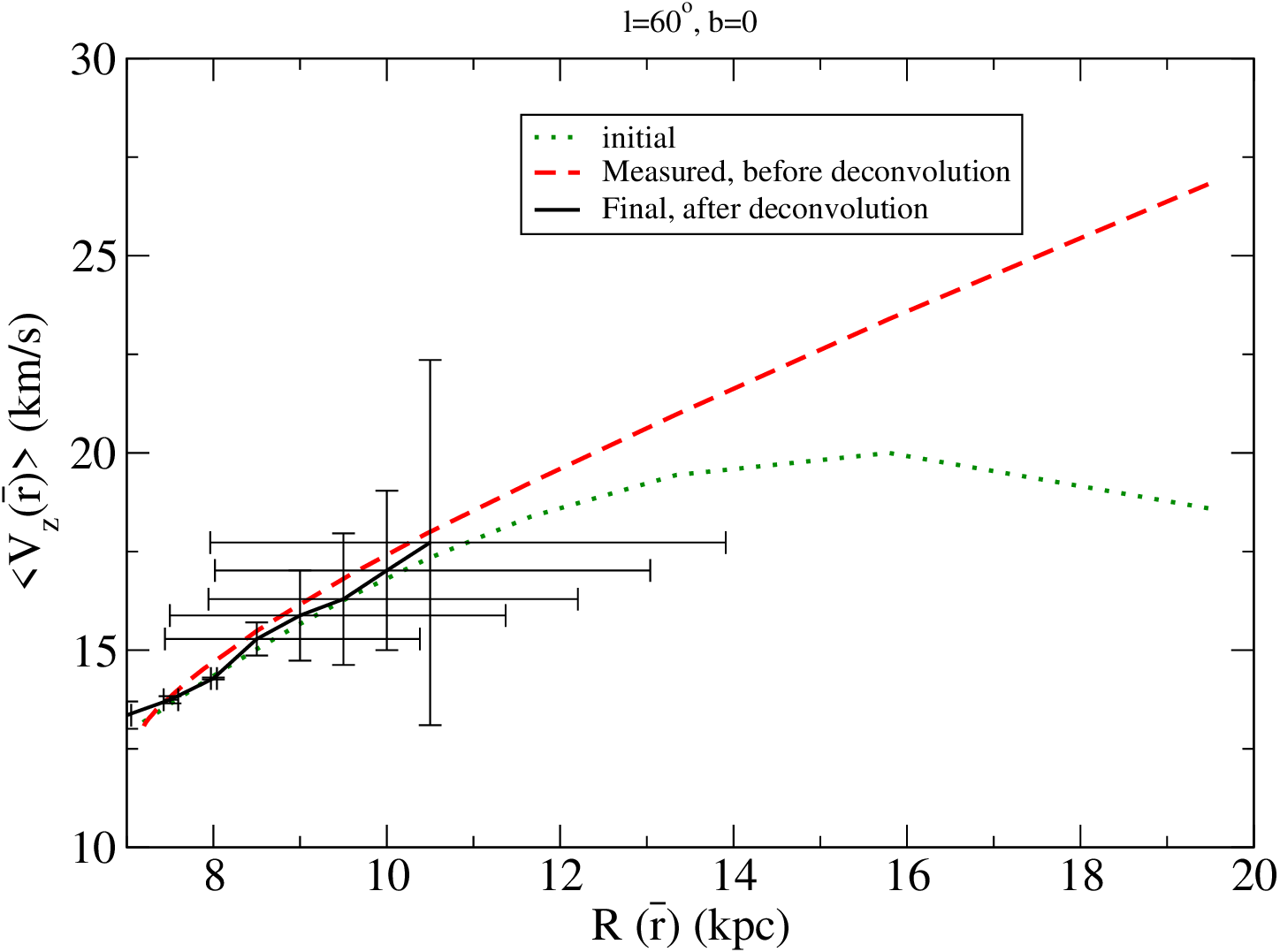}
\caption{Monte Carlo simulation of $V_z$. Horizontal error bars indicate the
r.m.s. value  of  the mean deconvolved $R$ (thus the indicated value is an average over that range of dispersion). The three panels are for the directions in the plane 
$\ell =180 ^\circ$, $\ell =120^\circ $ and $\ell =60^\circ $, respectively.}
\vspace{.2cm}
\label{Fig:simul4}
\end{figure}

In 
Figs. \ref{Fig:simul1}-\ref{Fig:simul4}
  we show 
 the results of our Monte Carlo simulations
 for the different components of the velocity and several
different directions in the plane.
We compare the ``initial'' assumed $\vec{V}$ with the one that we measure directly ``before deconvolution'', which departs very significantly from the initial $\vec{V}$, and the ``final'' result once we apply the deconvolution. The horizontal error bars indicate the
r.m.s. value  of  the mean deconvolved $R$ (thus the indicated value is an average over that range of dispersion). 
We see that, within the 
error bars, 
the recovered value of $\vec{V}$ agrees with 
the initial one, except for $V_\phi $ along $\ell =60^\circ $ because 
the uncertainties in the radial Galactocentric distance are so large that a large 
uncertainty in the separation of radial and azimuthal components along this line of sight is
produced. We see below
that these lines of sight are not useful, precisely because of that. 
We also note that, while the direct measurement of $R$ gives values up to 30 kpc in the anticenter, the corrected deconvolved values of $R$ are $\lesssim 18$ kpc. As expected, the large errors in parallax for distant sources is placing some stars beyond what can be observed, but the correction
adequately reduces the range.

The bias of differences within the noise between ``final'' and ``initial'' velocities  depends on the function and the direction, meaning that there is not always  positive or negative systematic error. Due to the asymmetry in the distribution (larger number of sources at lower heliocentric distances than farther away), the obtained values of velocities in the bins for large $R$ tend to give values at slightly smaller  $R$. One could try to correct this effect by assuming a priori the density distribution of stars, but we avoid the introduction of any prior in our method. We note that the determination of $N(\pi)$ will not change if we change the velocity field; however, the Lucy's method introduces a bias that can affect the calculation of $\vec{V}(\overline{r})$: the effect of such a  bias depends on the velocity field itself, as we observe here.

\section{Extended region kinematics}
\label{.results}

\begin{figure*}
\vspace{1.5cm}
\centering
\includegraphics[width=18cm]{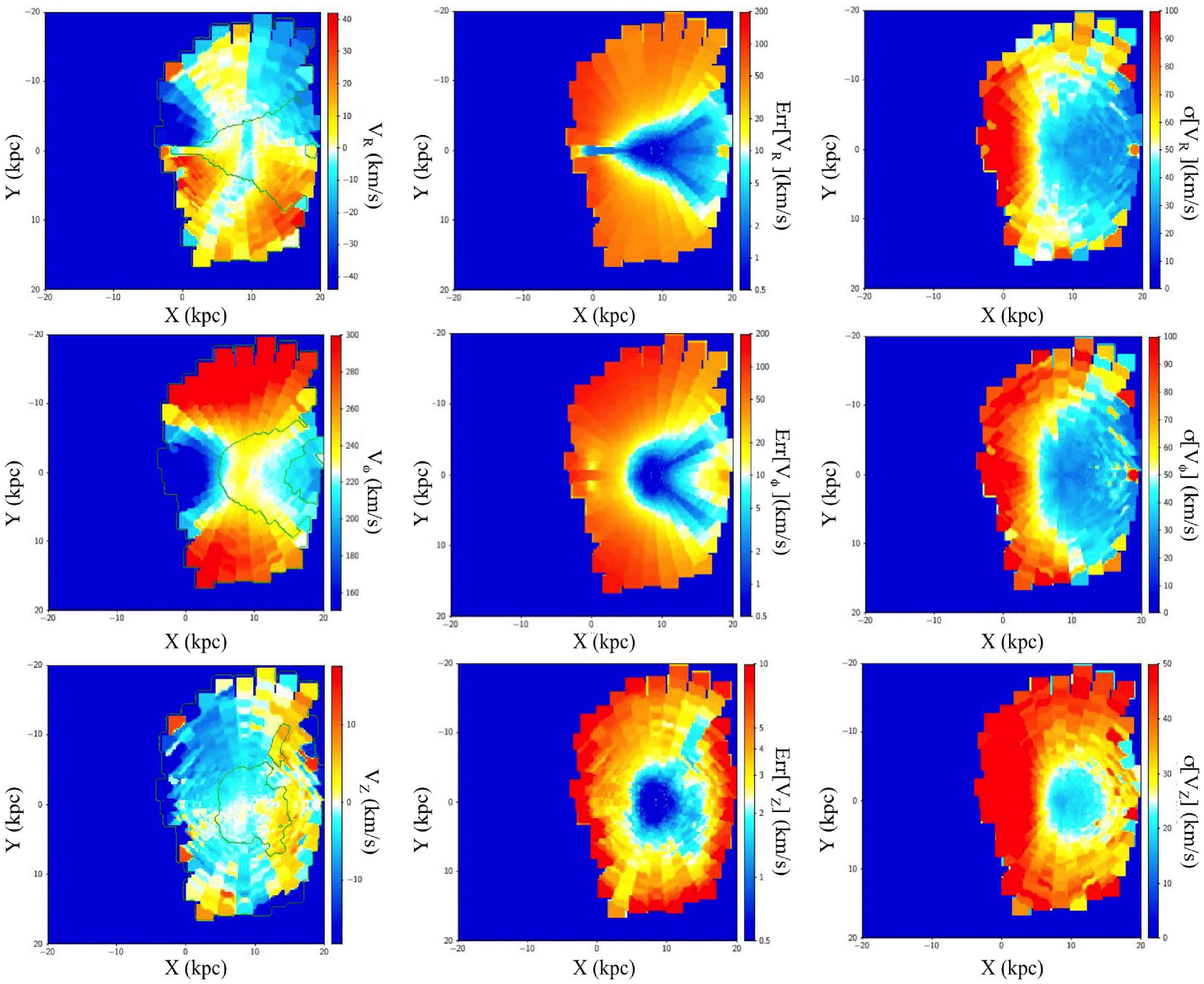}
\caption{
Left panels: The three components (radial, azimuthal, and vertical, from top to bottom, respectively) of the
Galactocentric velocity Median($\vec{V}$) as a function of $X(\overline{r})$ and
$Y(\overline{r})$ ($X$, $Y$ with deconvolution correction according to Sect. \ref{.deconv}).
Application for Gaia-DR2 data including $v_r$ measurement with the constraints:
$|b|<10^\circ $, $\frac{\pi}{\Delta \pi}>1$. Inner 
green contour represents the region
with error equal to 10, 10, and 2 km/s for $V_R$, $V_\phi, $ and $V_Z,$ respectively. 
Middle panels: mean error (in log scale) of $\vec{V}$ derived from the expansion
in the formulae of $\Delta \overline{r}=\sigma _{\overline{r}}$ as systematic
error and the dispersion of values of $v_r$, $\mu _\ell,$ and $\mu _b$
divided by the root square of the number of points per bin as statistical errors. 
 Right panels: r.m.s. value  of $\vec{V}$ (corrected for measurement errors).
The data used to make these plots are publicly available in the files fig8\_VR, fig8\_Vphi and fig8\_VZ of the URL 
www.iac.es/galeria/martinlc/codes/GaiaDR2extkin/.}

\vspace{.2cm}
\label{Fig:lucy2}
\end{figure*}

\begin{figure*}
\vspace{1.5cm}
\centering
\includegraphics[width=18cm]{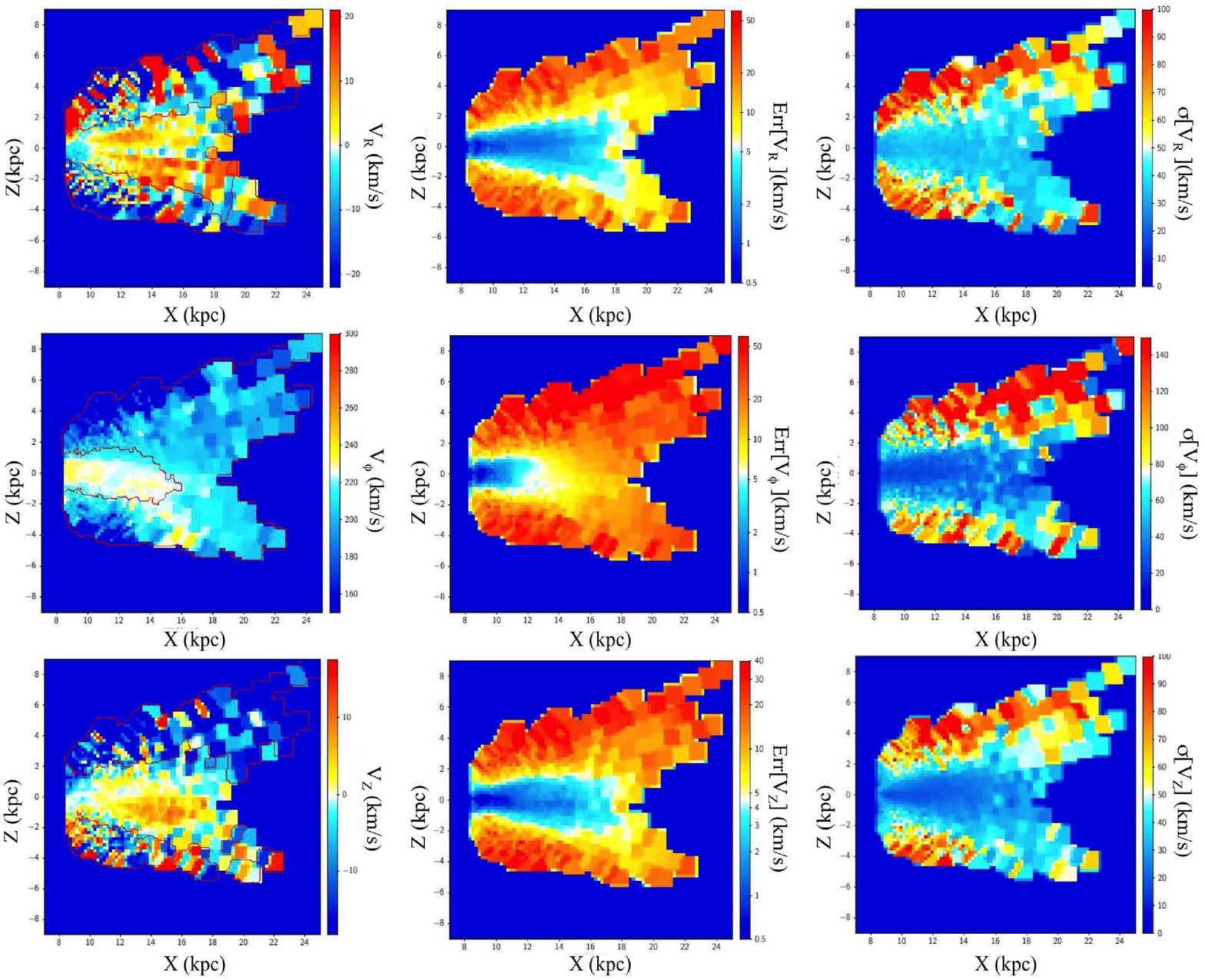}
\caption{Left panels: The three components (radial, azimuthal and vertical, from top to bottom, respectively) 
of the
Galactocentric velocity median ($\vec{V}$) as a function of $R=X(\overline{r})$ and
$Z(\overline{r})$ ($X$, $Z$ 
with deconvolution correction according to Sect. \ref{.deconv}).
Application for Gaia-DR2 data including $v_r$ measurement with the constraints:
$160^\circ <\ell <200^\circ $, $\frac{\pi}{\Delta \pi}>1$.
Inner red contour represents the region
with error equal to 10 km/s.
Middle panels: mean error (in log scale) of $\vec{V}$ derived from the expansion
in the formulae of $\Delta \overline{r}=\sigma _{\overline{r}}$ as systematic
error and the dispersion of values of $v_r$, $\mu _\ell,$ and $\mu _b$
divided by the root square of the number of points per bin as statistical errors. 
 Right panels: r.m.s. value  
 of $\vec{V}$ (corrected for measurement errors).
The data used to make these plots are publicly available in the files fig9\_VR, fig9\_Vphi and fig9\_VZ of the URL 
www.iac.es/galeria/martinlc/codes/GaiaDR2extkin/.}
\vspace{.2cm}
\label{Fig:lucy3}
\end{figure*}

\begin{figure*}
\vspace{1.5cm}
\centering
\includegraphics[width=18cm]{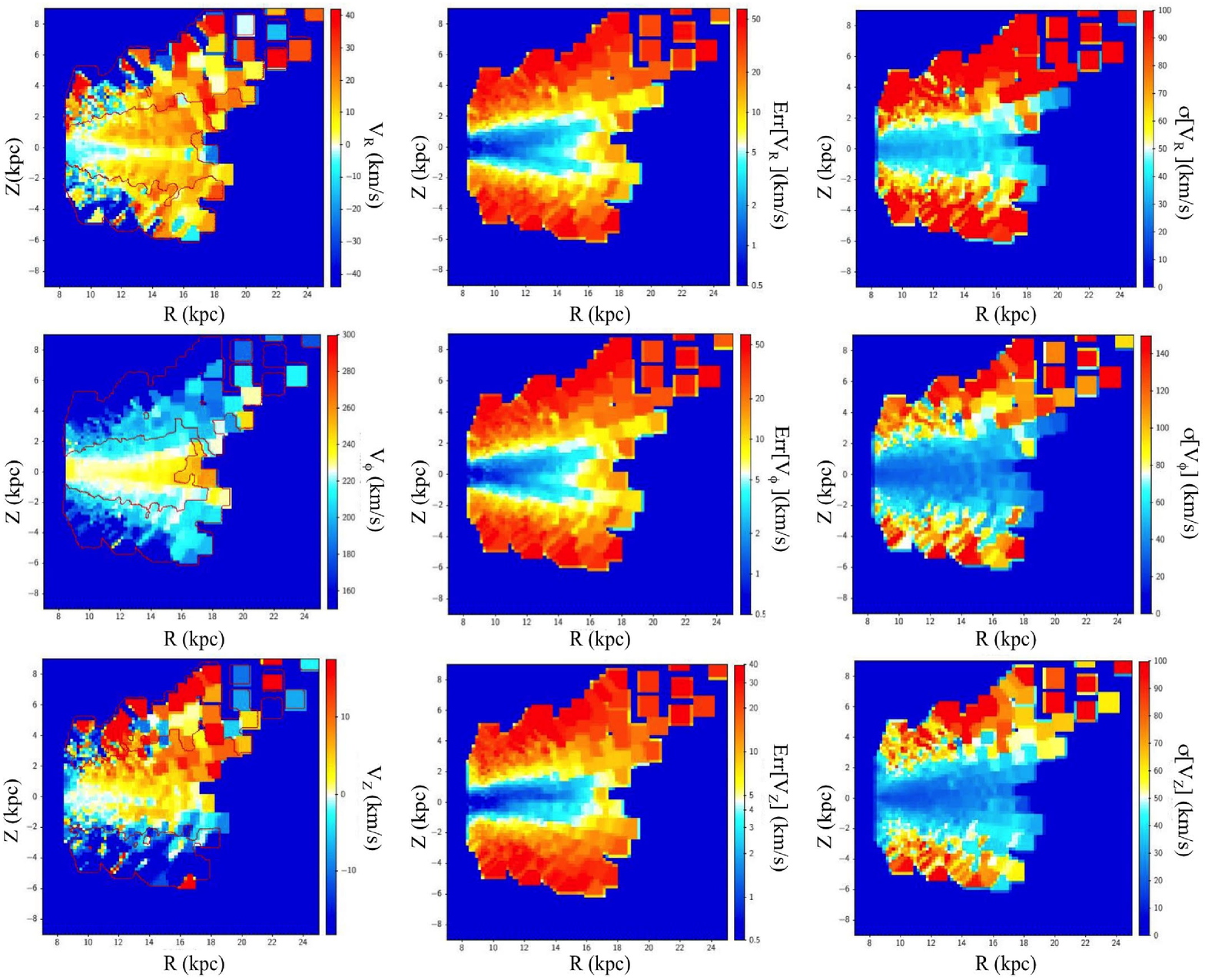}
\caption{As in Fig. \ref{Fig:lucy3} ($\vec{V}(R,Z)$, its error and r.m.s value) 
for $120^\circ <\ell <160^\circ $.
The data used to make these plots are publicly available in the files fig10\_VR, fig10\_Vphi and fig10\_VZ of the URL 
www.iac.es/galeria/martinlc/codes/GaiaDR2extkin/.}
\vspace{.2cm}
\label{Fig:lucy4_140}
\end{figure*}

\begin{figure*}
\vspace{1.5cm}
\centering
\includegraphics[width=18cm]{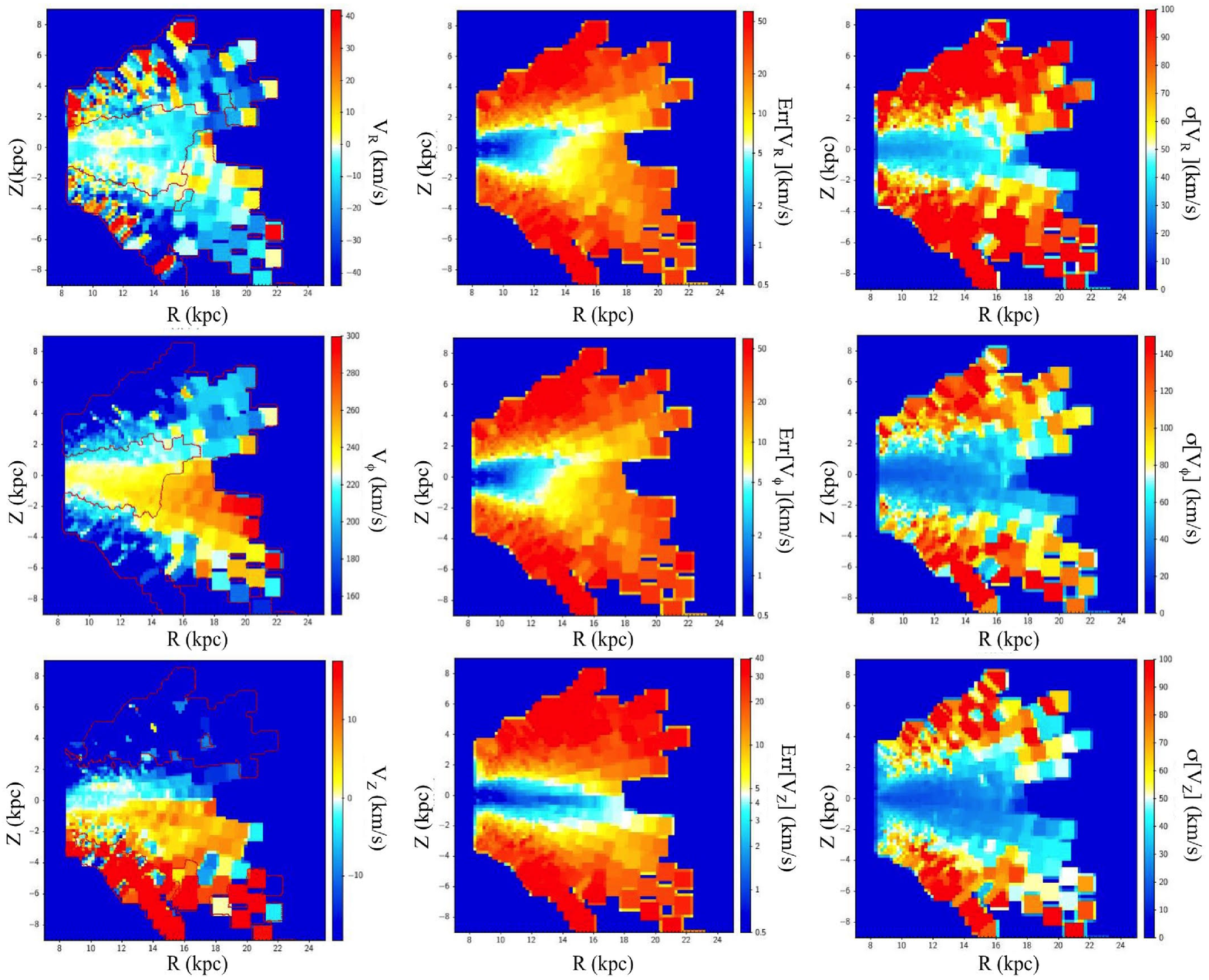}
\caption{As in Fig. \ref{Fig:lucy3} ($\vec{V}(R,Z)$, its error and r.m.s. value) 
for $200^\circ <\ell <240^\circ $.
The data used to make these plots are publicly available in the files fig11\_VR, fig11\_Vphi and fig11\_VZ of the URL 
www.iac.es/galeria/martinlc/codes/GaiaDR2extkin/.}
\vspace{.2cm}
\label{Fig:lucy4_220}
\end{figure*}

\begin{figure*}
\vspace{1.5cm}
\centering
\includegraphics[width=18cm]{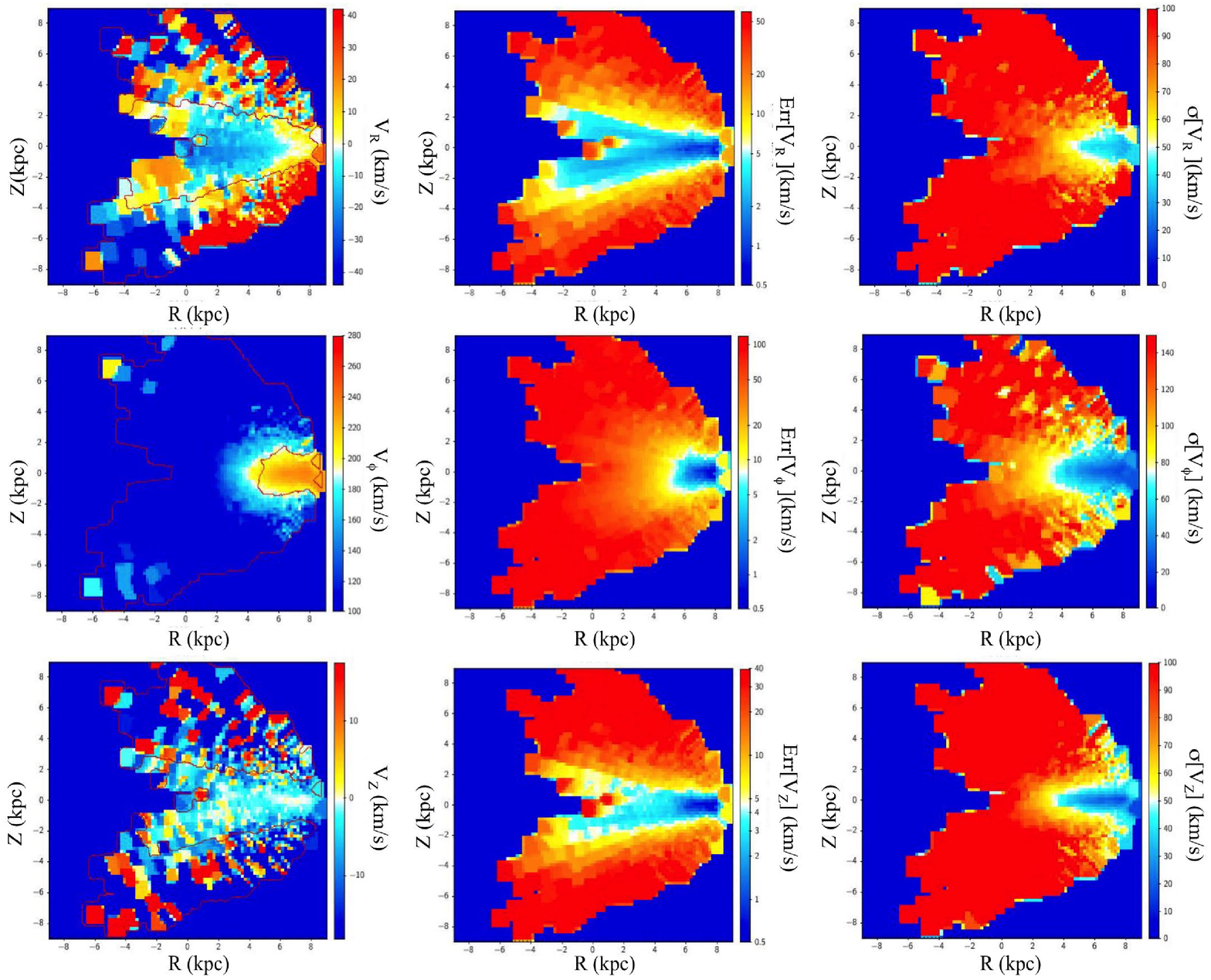}
\caption{As in Fig. \ref{Fig:lucy3} ($\vec{V}(R,Z)$, its error and r.m.s. value) 
for $-20^\circ <\ell <+20^\circ $.
The data used to make these plots are publicly available in the files fig12\_VR, fig12\_Vphi and fig12\_VZ of the URL 
www.iac.es/galeria/martinlc/codes/GaiaDR2extkin/.}
\vspace{.2cm}
\label{Fig:lucy4_0}
\end{figure*}

We can now apply the above method to obtain the median $\vec{V}$.
In Figs. \ref{Fig:lucy2}-\ref{Fig:lucy4_0}
we plot the results in the Galactic 
plane of all stars with $\Delta \pi <\pi$ and the following constraints on coordinates:
\begin{enumerate}
\item $|b|<10^\circ $: this gives a wider range in $Z$ for larger
 distance from the Sun, while in the solar neighbourhood ($\overline{r}\sim 1$ kpc) it is constrained to $Z\lesssim 0.2$ kpc, and at 
$\overline{r}\sim 10$ kpc it is constrained to  $Z\lesssim 2$ kpc. This is enough to sample most disk stars, including the effect of the flare in the outermost disk. Certainly, in the center of the Galaxy ($R\lesssim 4$ kpc) we are overexploring the region of the disk and we are also including the bulge component. The sky within the constraint was divided into 36 lines of sight, each of them
with $\Delta \ell=10^\circ $, and in each of these cells 
 we have applied the above deconvolution technique.
\item $160^\circ <\ell <200^\circ $:
in the anticenter region, which is a region where velocity errors are the lowest, 
we explore the dependence on vertical distance.
The sky was divided into 36 lines of sight, each of them
with $\Delta b=5^\circ $, and for those areas we have applied the above deconvolution technique.
\item  As in 2. but for $120^\circ <\ell <160^\circ $.
\item  As in 2. but for $200^\circ <\ell <240^\circ $.
\item  As in 2. but for $-20^\circ <\ell <20^\circ $.
\end{enumerate}

In all figures, only the bins with number of stars $N\ge 6$ are plotted.
For the mean error on $\vec{V}$, we assume,  
from the propagation
of the  errors,
that 
\begin{equation}
\Delta \overline{r}=\sigma _{\overline{r}}
;\end{equation}
this  is a systematic
error, so it cannot be reduced by the increase of the 
number of stars
 $N$ in each bin.
In addition, we include in the propagation of errors,
\begin{equation}
\Delta v_r=\frac{\sigma (v_r)}{\sqrt{N}} \;; 
\Delta \mu _\ell= \frac{\sigma (\mu _\ell )}{\sqrt{N}}
\;\; \mbox{and} \; \; 
\Delta \mu _b =\frac{\sigma (\mu _b)}{\sqrt{N}},\end{equation}
so these are interpreted as statistical errors that can be  reduced
by increasing the number of stars. 
Here we neglect the covariance terms in the errors on
 $\mu _\ell$, $\mu _b,$ and $\overline{r}$: that is,
we assume these errors are independent from each other and thus we neglect
that these 
quantities are not derived in isolation
and that there may be some correlation between them 
(Luri et al. 2018, Sect,  2.2).
The r.m.s. values plotted in the right panels of
Fig. \ref{Fig:lucy2} were corrected (subtracted quadratically) from the
measurement errors in $v_r$, $\mu _\ell, $ and $\mu _b$; however, we have not taken
into account the variations of $\vec{V}$ due to the dispersion of $r$ with the deconvolution
(i.e., we neglect $\nabla \vec{V}$) and there may also remain
some residuals in high-error regions.
In Fig. \ref{Fig:lucy2}, uncertainties on $V_R$ and $V_\phi $ 
are smaller towards the anticenter because  the separation of both components
 is independent of the distance. 
Moreover, $V_R$ only depends on $v_r$, so it is insensitive to the
errors on the distance for that reason too.
In 
Figs. \ref{Fig:lucy3}-\ref{Fig:lucy4_0}, 
there are larger uncertainties towards the Galactic pole due 
to the low number of sources with our binning with 
constant $\Delta b $.

The results for $\overline{r}<4$ kpc are very similar to those of Fig. 10 (for $|z|<0.2$ kpc)
and  Fig. 11  (for $|\phi |<15^\circ $) of G18 with the same Gaia-DR2 data: this 
corroborates the reliability of our analysis in the common range of distances. 
 However, as we discuss in the following subsections, 
the most interesting features are observed
at larger heliocentric distances, where our analyses provide a 
novel contribution.

\subsection{Radial velocities}
The median radial velocities, their errors, and the r.m.s. values 
are plotted 
in the three top panels of 
Figs. \ref{Fig:lucy2}-\ref{Fig:lucy4_0} respectively.

\begin{figure*}
\vspace{1.5cm}
\centering
\includegraphics[width=12cm]{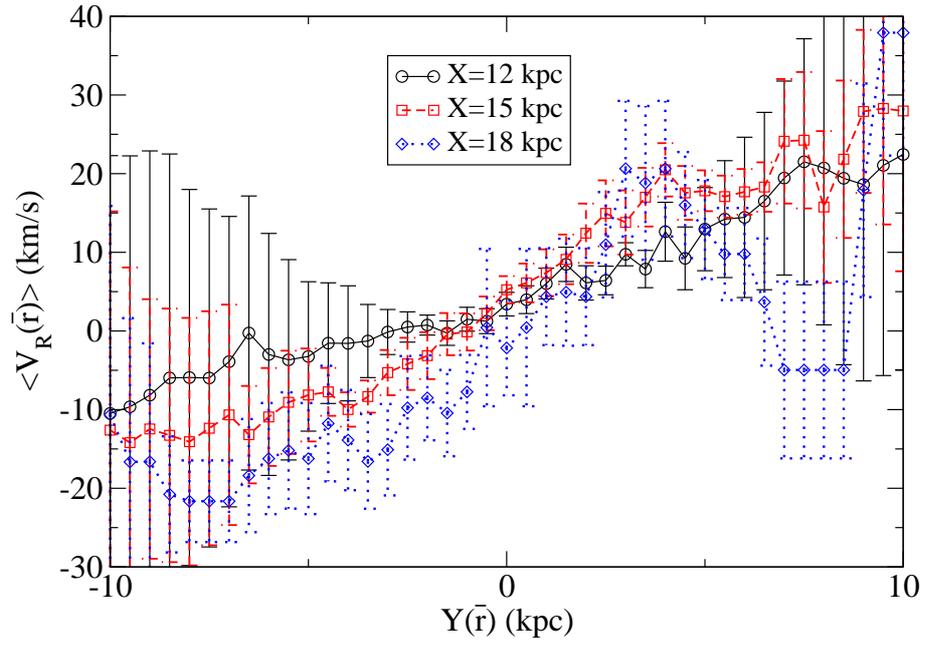}
\caption{ Radial
Galactocentric velocity median as a function of $Y(\overline{r})$ for different $X(\overline{r})$ ($\Delta X=0.5$ kpc) within $|b|<10^\circ $.}
\vspace{.2cm}
\label{Fig:vry}
\end{figure*}

Regions in the top-middle panel of Fig. \ref{Fig:lucy2} with yellow-red colors indicate the areas with errors larger than $\sim 10$ km/s, thus indicating where the median velocities are not accurate. The regions 
in  white-blue tones have  errors lower than 10 km/s,  darker
blue being indicative of errors lower than 1 km/s. In the $X-Y$ plane, a prominent feature within low-error regions ($\Delta V_R\lesssim 10$ km/s)
is represented by 
the gradient of velocities along the $Y$ direction 
with  $X>10$ kpc:
between positive median values of $V_R\sim +20$ km/s at $X>12$ kpc and $2<Y({\rm kpc})<10$ 
(in orange color in top-left panel of Fig.  \ref{Fig:lucy2})
and $V_R\sim -20$ km/s at $X>16$ kpc and $-8<Y({\rm kpc})<0$
(in medium blue in top-left panel of Fig.  \ref{Fig:lucy2}).
Hence, we have a scenario of expansion (stars moving outwards) 
in the second quadrant and contraction (stars moving inwards) in the third quadrant, and a
transition between both areas in a line approximately between
($X=10,Y=-5$) and 
($X=18,Y=0$) with a radial velocity gradient between these two regions
of about 40 km/s. This is also shown in  Fig. \ref{Fig:vry}.

Furthermore, the variation along the  center-Sun-anticenter direction shows a change of sign of the radial velocity:
positive 
for $X<8$ kpc, negative for $8<X({\rm kpc})<10$ and positive for $10<X({\rm kpc})<16$ kpc. This is also observed in the top panel of Fig. \ref{Fig:vr}.

This same behavior has already been noted for regions with $R\lesssim 13$ kpc (Siebert et al. 2011; Williams et al. 2013; Wang et al. 2018; Carrillo et al. 2018; G18), and there are even works that reached $R\approx 16$ kpc (L\'opez-Corredoira \& Gonz\'alez-Fern\'andez 2016; Tian et al. 2017). 

By examining the dependence with $Z$ in the regions 
where the errors are
lower than 5 km/s (i.e., the blue regions of the top-middle panels
of  Fig. \ref{Fig:lucy3}-\ref{Fig:lucy4_220}) 
we may notice 
that there is more contribution from these positive velocities in the anticenter (Fig.
\ref{Fig:lucy3}) at $X>10$ kpc and $Z<0$ than at $Z>0$, but not in other azimuths. This is also similar to the results in Wang et al. (2018) only for $R<13$ kpc.
We think that the special feature of the anticenter with respect to this asymmetry is
real, but it has nothing to do with systematic effects towards that line of sight
as we could not identify any source of systematic error in the data. The fact that this 
is found at $\phi \approx 0$ and not in other azimuths is therefore a coincidence.

The values of $\sigma (V_R)$ plotted in the top-right panel of Figs. \ref{Fig:lucy2}-\ref{Fig:lucy4_0} show no abnormality: $\sigma (V_R)$ is higher for lower $R$ or larger $|Z|$. The values and the trends are in agreement with previous estimations: for example, Williams et al. (2013), G18.
For $\ell=0$ (Fig. \ref{Fig:lucy4_0}: middle-right),
$\sigma (V_\phi)$ is dominated by large  uncertainties towards the Galactic center.


\subsection{Azimuthal velocities}

The median azimuthal velocities, their error, and the r.m.s.  values are plotted 
in the three middle row panels, from left to right, respectively, of Figs. \ref{Fig:lucy2}-\ref{Fig:lucy4_0}.

Regions in the middle panel of Fig. \ref{Fig:lucy2} with yellow, orange, and red colors indicate the areas with errors larger than $\sim 10$ km/s, thus 
highlighting where the median velocities are inaccurate. The regions in  white and light-blue tones have errors lower than 10 km/s, the darker
blue being  indicative of errors lower than 1 km/s. 
In the $X-Y$
plane, within low-error regions ($\Delta V_\phi \lesssim 10$ km/s), 
apart from some
small fluctuations  
there is a clear decrease of the
 azimuthal velocity for $R>10$ kpc:
that is, we find\footnote{Note that $V_\phi (R_\odot )\ne V_{g,\odot }$ because the first term describes the azimuthal motion of the Local Standard of Rest [LSR], whereas the second term also includes the motion of the Sun with respect to the LSR.}  $V_\phi \lesssim 220$
km/s for $R>12$ kpc, but $V_\phi (R_\odot )\approx 240$ km/s. 
This is also observed in the middle panel of Fig. \ref{Fig:vr}. We note however that higher $R$ also means higher average $|Z|$ (since we have a constant range of Galactic latitude), so part of this decrease is
due to a larger ratio of high $|Z|$ stars.

Figure \ref{Fig:lucy3} (left column, middle row) provides a more surprising insight into the distribution of
azimuthal velocities towards 
the anticenter. If we pay attention only to low-error pixels ($\Delta V_R\lesssim 10$ km/s, that is, excluding the regions in orange and red in the central panel of Fig. \ref{Fig:lucy3}), we see a clear decrease of velocity for larger $|Z|$, and, moreover, we see an asymmetry by which the velocity is higher for $Z<0$ than for $Z>0$ in the range $12<R({\rm kpc})<16$.

\begin{figure*}
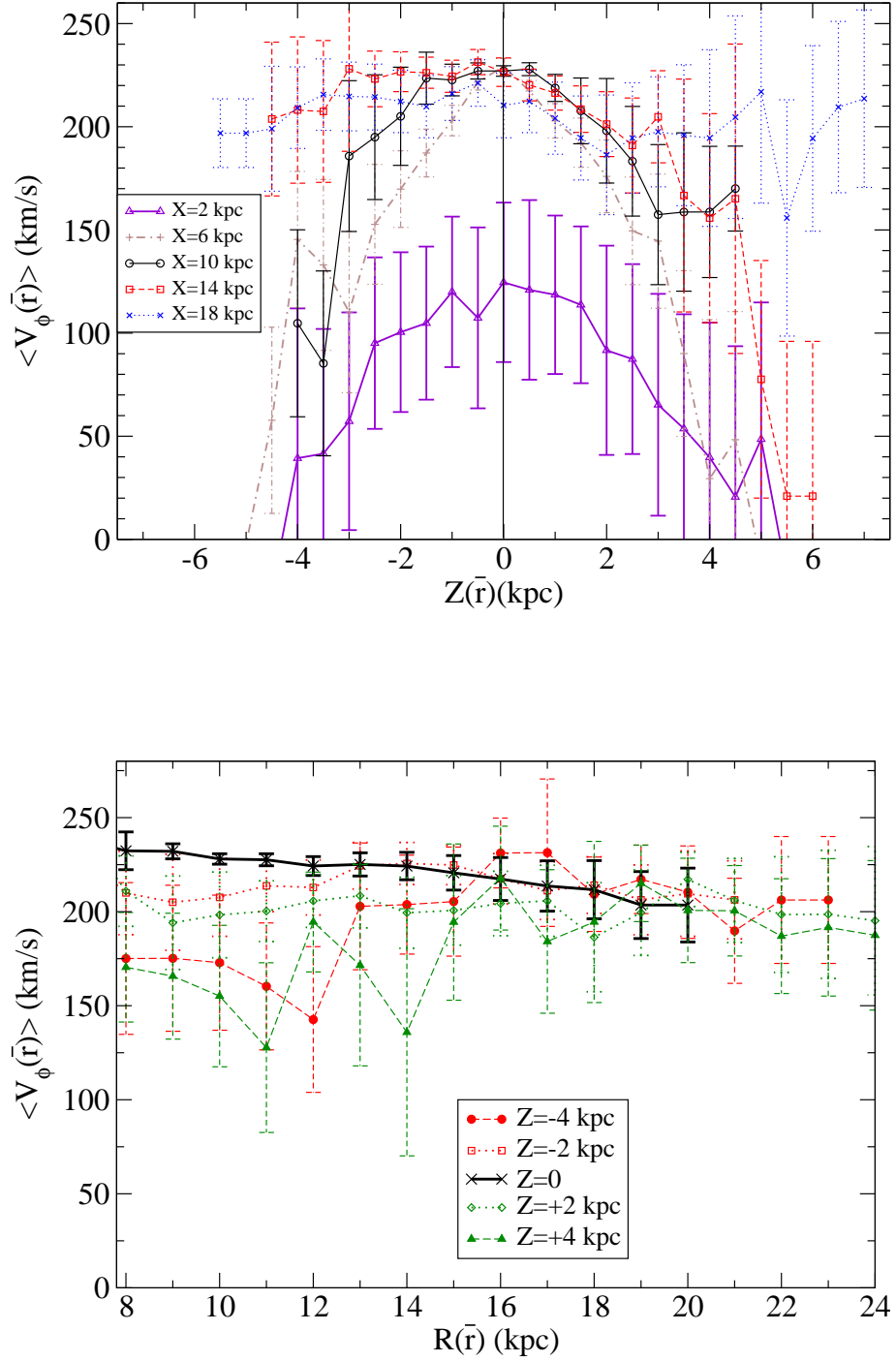

\vspace{1.5cm}
\centering
\includegraphics[width=12cm]{vphiz.eps}\\
\vspace{2cm}
\includegraphics[width=12cm]{rotvel.eps}
\caption{Azimuthal 
Galactocentric velocity median as a function of $Z(\overline{r})$ for different cuts of constant and $R=X(\overline{r})$ ($\Delta X=0.2$ kpc) (top) or as a function of $R=X(\overline{r})$
for different cuts of $Z(\overline{r})$ (bottom).
Application for Gaia-DR2 data including $v_r$ measurement with the constraints:
$-20^\circ <\ell <20^\circ $ (only top) or
$160^\circ <\ell <200^\circ $, $\frac{\pi}{\Delta \pi}>1$.}
\vspace{.2cm}
\label{Fig:vphiz}
\end{figure*}

In the top part of Fig. \ref{Fig:vphiz}, we show 
$\langle V_\phi (r) \rangle $
plots for different $Z$ values with
 the constraints  $|\ell -180^\circ |<20^\circ $ and 
  $|\ell |<20^\circ $, and observe the
same trend. In general for low $R<12$ kpc, we see a symmetric distribution with respect to the
midplane decreasing with $|Z|$; for $12<R({\rm kpc})<16$, there is a decrease of velocity
only for $Z>0$ while for negative $Z$ it remains flat and starts to decrease for $Z\lesssim -3.5$ kpc; and for $R>16$ kpc it remains almost flat without decrease with $Z$.
However, again this asymmetry is only observed 
towards the anticenter direction, but not along 
the lines of sight $\ell =140^\circ $, $\ell =220^\circ, $
and $\ell =0$ 
(Figs. \ref{Fig:lucy4_140}-\ref{Fig:lucy4_0}: left column, middle row).
We also highlight the radial gradient in Fig. \ref{Fig:vphiz}, a trend
already noted in Fig. \ref{Fig:vr} (middle) which represents 
the integral in $Z$ within the limits of $|b|<10$ deg. of Fig. 
\ref{Fig:vphiz}.

The monotonous decrease for 
$R\gtrsim 12$ kpc in the plane ($Z=0$) 
has already been found in multiple works for the rotation curve
(e.g., Dias \& L\'epine 2005; Bovy et al.
2012; Kafle et al. 2012; Reid et al. 2014; 
L\'opez-Corredoira 2014; Galazutdinov et al. 2015; G18), 
although we must bear in mind that an asymmetric drift correction (Bovy et al. 2012; 
Golubov et al. 2013) remains to be applied and some part of this decrease might be 
attributed to this correction.
The asymmetry north-south has not gained as much attention,  although L\'opez-Corredoira (2014), for example, already noted its existence for $R<16$ kpc, similarly to Wang et al. (2018) and Kawata et al. (2018)
for $R=11-12$ kpc. Furthermore, careful examination of Fig. 13(left) of G18 reveals some asymmetry at the maximum reached, $R\approx 13$ kpc.
A decrease of the azimuthal velocity with $|Z|$ was also previously known, although
with less-precise data and studying a lower $R$ range than here (e.g., Bond et al.
2010; Bobylev \& Bajkova 2010; 
Williams et al. 2013; L\'opez-Corredoira 2014; Wang et al. 2018; G18).
The extremely low value given by L\'opez-Corredoira (2014) of 
$V_\phi =82\pm 5$(stat.)$\pm 58$(syst.) km/s at $R=16$ kpc and $Z=2$ kpc
dominated by the anticenter stars is not confirmed here to be as low, but we get $V_\phi=204.8\pm 17.1$ km/s for that position towards the anticenter;
we note however that these values are compatible within $2\sigma $.

Like in the case of radial velocities, the values of $\sigma (V_\phi)$ plotted in the middle-right panel of Figs. \ref{Fig:lucy2}-\ref{Fig:lucy4_220} do not reveal any surprises either: $\sigma (V_\phi )$ is higher for lower $R$ or larger $|Z|$. For $\ell=0$ (Fig. \ref{Fig:lucy4_0}: middle-right)
$\sigma (V_\phi)$ is dominated by large
 uncertainties towards the Galactic center.
The values and the trends are in agreement with previous estimations (e.g., Williams et al. 2013, G18), although extending here to
larger values of $R$. We note that thin and thick disks are  included here and
show differing rotation dispersion between them (Haywood et al. 2013).


\subsection{Vertical velocities}

The median vertical velocities, their error and the r.m.s. values 
are plotted respectively 
in the three bottom panels of 
Figs. \ref{Fig:lucy2}-\ref{Fig:lucy4_0}.

\begin{figure*}
\vspace{1.5cm}
\centering
\includegraphics[width=12cm]{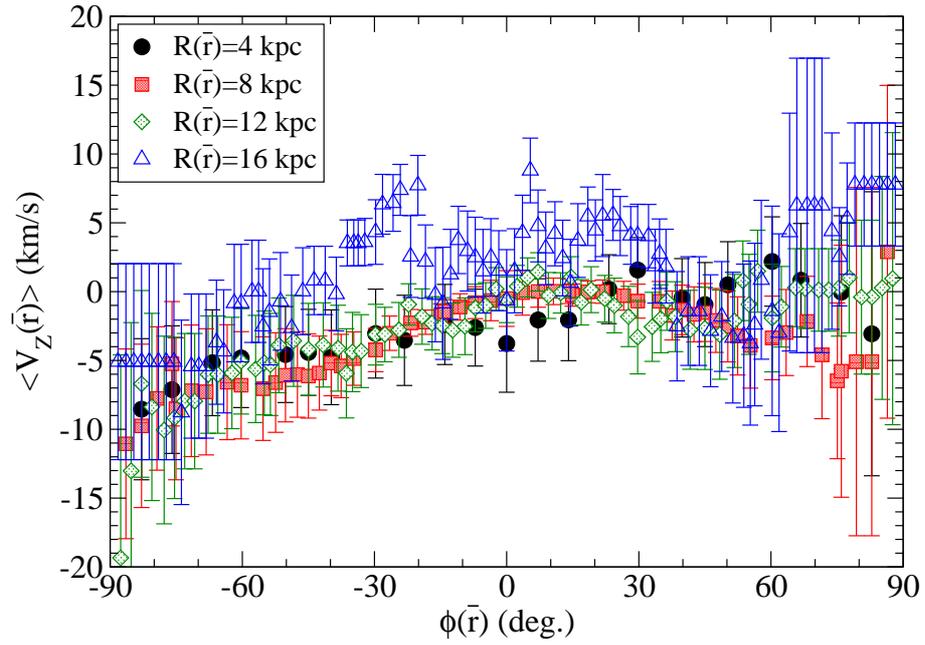}
\caption{Vertical
Galactocentric velocity median as a function of $\phi (\overline{r})$ for different Galactocentric radii ($\Delta R=0.5$ kpc) within $|b|<10^\circ $.}
\vspace{.2cm}
\label{Fig:vzphi}
\end{figure*}

Regions in the bottom-middle panel of Fig. \ref{Fig:lucy2} with white, yellow, and red colors indicate the areas with errors larger than $\sim 2.0$ km/s. The regions 
 with blue tones have  errors lower than 2.0 km/s, with darker blue being indicative of errors lower than 1.0 km/s. 
In the
$X-Y$ plane, a prominent feature within significant detection 
regions (${|V_Z|}/{\Delta V_Z}>3$) is the gradient of velocities approximately along the $X$ direction: between a minimum negative median value of $V_Z=-3.5\pm 1.1$ km/s at ($X=9.4$, $Y=-2.8$)  and $V_Z=+8.8\pm 2.4$ km/s at ($X=15.6$,
$Y=1.8$). In general, there is a trend of positive $V_Z$ for $R\gtrsim 12$ kpc and 
negative $V_Z$ for $R\lesssim 12$ kpc.
As shown in Fig. \ref{Fig:vzphi},
the gradient is also present with respect to the azimuth.
In this figure, we also observe that
the radial gradient in  $V_Z$ 
 becomes significant when comparing the green 
 diamonds ($R=12$ kpc) and the blue triangles ($R=16$ kpc). The bottom row of
Fig. \ref{Fig:lucy3}
provides evidence that a larger contribution to 
positive $V_Z$, with values even larger than +10 km/s, 
comes from the southern hemisphere
in the anticenter direction:   the
north-south asymmetry is clear. 
For other lines of sight different from the anticenter, we also see asymmetries, but
larger in amplitude for $\ell =220^\circ $ (Fig. \ref{Fig:lucy4_220}: bottom), and
with opposite sign for $\ell =140^\circ $ (Fig. \ref{Fig:lucy4_140}: bottom).
Towards the center (Fig. \ref{Fig:lucy4_0}: bottom), no asymmetry is observed.

L\'opez-Corredoira et al. (2014) in a sample within the range $5<R({\rm kpc})<16$
noticed both the north-south  asymmetry and the trend that 
 $V_Z$ increases with $R$, 
 but found a different absolute value of these velocities and their 
azimuthal dependence. This  is not surprising as the data considered by 
L\'opez-Corredoira et al. (2014)  were very
noisy and they only got a detection within 2-$\sigma $, which we could interpret as a random fluctuation.
Liu et al. (2017), Wang et al. (2018), Kawata et al. (2018), and G18 also noted 
 both the  north-south asymmetry and the increase of $V_Z$ with $R$, but measured only up to $R=13-14$ kpc. 
Liu et al. (2017) indeed observed that stars with ages $<2$ Gyr present a 
significantly a positive vertical velocity in the anticenter higher than the red clumps with age $>2$ Gyr. Also, Widrow et al. (2012), Williams et al. (2013), and Carrillo et al. (2018) observed some of these features but only for regions within 2 kpc of the Sun.
Finally, it is worth mentioning that Poggio et al. (2018) observed similar maps using Gaia+2MASS out to $R=15$ kpc.

Like in the case of radial velocities, the values of $\sigma (V_Z)$ plotted in the bottom-right panel of Figs. \ref{Fig:lucy2}-\ref{Fig:lucy4_220} do not reveal any surprises either: $\sigma (V_Z)$ is higher for lower $R$ or larger $|Z|$, and in this case also for larger $r$ associated to larger values of average $|b|$, explaining the red ring in the bottom-right panel of Fig. \ref{Fig:lucy2}. The values and the trends are in agreement with previous estimations (e.g., Williams et al. 2013, G18), although extending here to larger values of $R$.
For $\ell=0$ (Fig. \ref{Fig:lucy4_0}/bottom-right)
$\sigma (V_Z)$ is dominated by large
 uncertainties towards the Galactic center.
We again note that thin and thick disks are included here and
display differing vertical dispersion between them (Haywood et al. 2013).

\subsection{Effect of a global zero-point bias in the parallaxes}
\label{.0point}

Up to now, we have not considered any possible zero-point bias in the parallaxes
of Gaia-DR2. The most accurate measurements indicate that there is
a systematic bias of -0.03 mas. (Lindegren et al. 2018, Sect. 5.2; Arenou et al. 2018, Sect. 4.4) in the 
sense that the Gaia-DR2 parallaxes are smaller than 
the true value, apart from statistical errors. 

\begin{figure*}
\vspace{1.5cm}
\centering
\includegraphics[width=18cm]{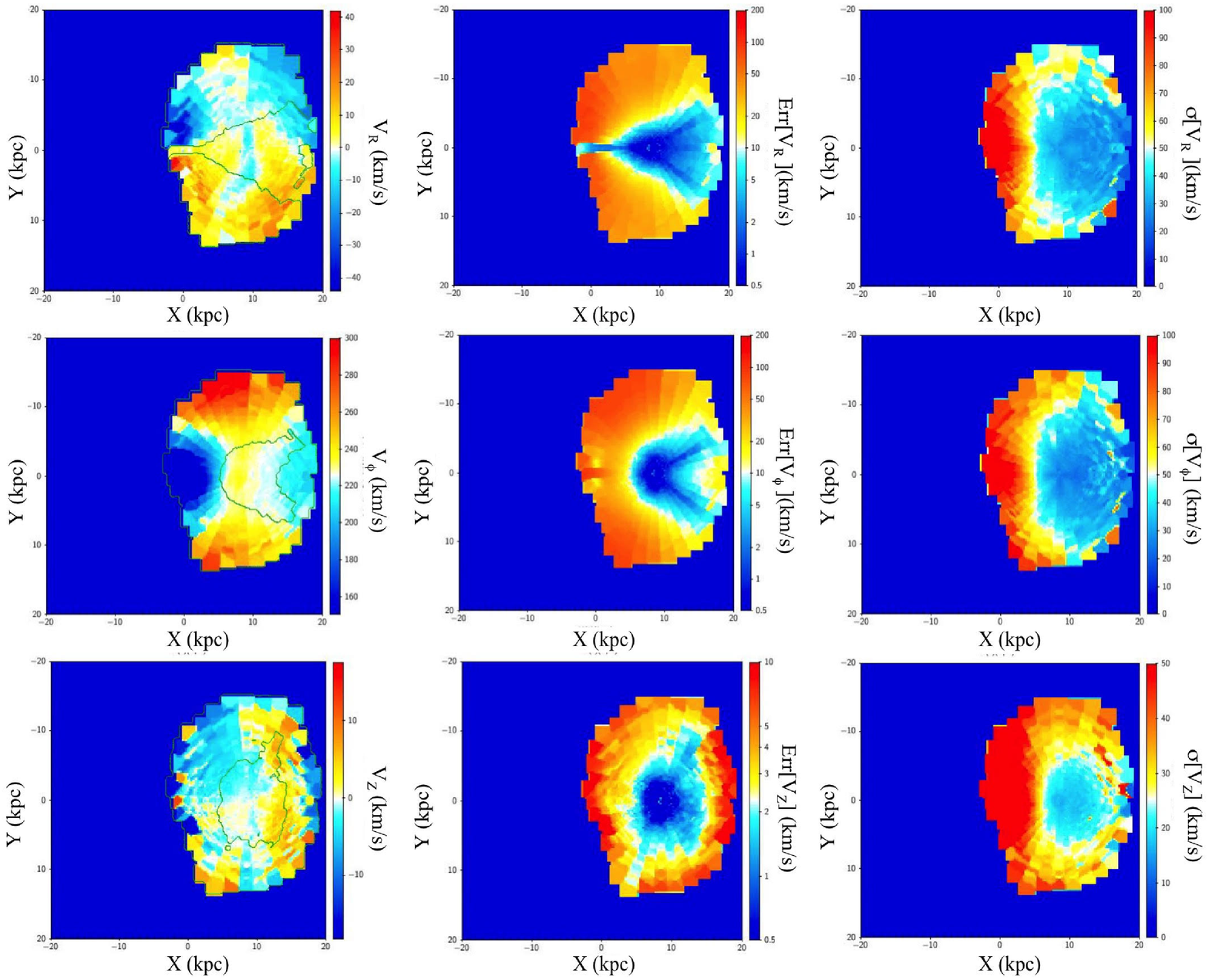}
\caption{As in Fig. \ref{Fig:lucy2}, this time introducing a correction to
the zero-point bias of parallaxes $\pi _c=\pi +0.03$ mas.
The data used to make these plots are publicly available in the files fig16\_VR, fig16\_Vphi and fig16\_VZ of the URL 
www.iac.es/galeria/martinlc/codes/GaiaDR2extkin/.}
\vspace{.2cm}
\label{Fig:lucy2_0point}
\end{figure*}

By repeating our calculations of Fig. \ref{Fig:lucy2} 
with a corrected parallax $\pi _c=\pi +0.03$ mas., we obtained the results shown in Fig.
\ref{Fig:lucy2_0point}. The results are very similar and deserve the same considerations
as those already described above for Fig. \ref{Fig:lucy2}. 
The only difference is the very large error bar pixels
that show a better trend with this correction, but these pixels were not taken into account
in the previous comments so the conclusions remain the same. 
A similar situation occurs also with the X-Z plots.
A higher zero-point bias was 
 also calculated by Stassun \& Torres (2018) or by 
 Zinn et al. (2018)
but it had the same order of magnitude, 
so the effect of its correction must be
of the same order.

\section{Discussion and conclusions}
\label{.disc}

An analysis on the dynamical interpretation of these extended kinematic
maps will be given in a forthcoming paper (Part II of this paper, in preparation). Nonetheless, without entering here into a discussion about
the possible theoretical scenarios, we provide a few notes and considerations from a purely observational point of view.

First, it is clear that 
our Galaxy is not 
an axisymmetric system at equilibrium 
where orbits are purely circular,  that is, with $V_R=V_Z=0$ and $V_\phi (R,\phi, Z)=V_\phi (R)$;  this is especially true  for the outer disk.
Many other authors  have 
recently reached this  
same conclusion (e.g., Antoja et al. 2018; G18; Kawata et al. 2018).
The presence of a radial and azimuthal velocity gradient of about $40$ km/s,
 of a vertical velocity gradient of 10 km/s, and of the
 north-south asymmetry, with in general higher speeds in the southern Galactic hemisphere, are remarkable characteristics that clearly indicate 
  that the disk of the Milky Way is not an
axisymmetric system at equilibrium, but that it is characterized by
streaming motions in all three velocity components.

In addition we confirm that the azimuthal velocity is not flat for $R>R_\odot $. 
Even taking account constant $Z$ strips (see
Fig. \ref{Fig:vphiz}) rather than constant $b$ cuts, we see some decrease
of the azimuthal velocity: at $R=18$ kpc it is $\approx 10$\% lower than at $R=R_\odot $. Other authors (e.g., Galazutdinov et al. 2015) have also noted
the fall-off of this velocity, which is more significant for $R>15$ kpc.
However,  here we have computed only the average velocity in some bins of stars
as a function of their position with respect to the center of the Galaxy, which is not strictly equal to the rotation speed and needs to be corrected by the
asymmetric drift, whose expected correction value is $\lesssim 20$ km/s in the Galactic plane 
(Bovy et al. 2012, Golubov et al. 2013) and larger for $Z\ne 0$ given the
higher values of velocity dispersions in off-plane regions (G18). It is possible that 
the total dependence of $V_\phi $ on $Z$, or at least part of it, may be explained 
by this effect. This is something that we will explore in Part II 
of this paper, together 
with other calculations related to dynamics and the Jeans equations 
(Kafle et al. 2012).

Vertical motions are small, constrained to be within an absolute value lower than 10 km/s, and present some patterns of oscillation and gradients and are different in general for
the northern and southern Galactic hemispheres. Other north-south 
asymmetries in $V_R$ or $V_\phi $ only show up in the
anticenter direction.

Future analyses of Gaia data
with the use of the deconvolution of parallax errors by
means of Lucy's method of inversion, which we discuss in this work, 
will allow the explored region to  $R$ to be increased far beyond 20 kpc,
both because future data releases of Gaia will provide a much deeper 
magnitude limit in the subsample including radial velocities, 
and because the parallax errors will also be much lower.
The range of distances covered by studies using only stars with distance errors
$<20$\% (such as the one that was done by G18) will be increased with the future data releases; 
however by considering the
deconvolution method we have discussed here, one can obtain
 information at heliocentric distances that are two to three times larger than results without 
applying error deconvolution techniques.
Furthermore, the present method can also be used to obtain the density $\rho (r)$ along
the line of sight once we know the luminosity function, 
even for larger error in parallaxes, by means of Eqs. (\ref{convol}) and
(\ref{Nselec}), 
which is another application that will allow us to extend our
knowledge of the morphology of the disk and of the 
stellar halo at very large distances.


\begin{acknowledgements}

Thanks are given to the anonymous referee for helpful suggestions, which improved this paper, and
to Joshua Neve (language editor of A\&A) for proof-reading of the text.
MLC was supported by the grant AYA2015-66506-P of the Spanish Ministry of Economy and Competitiveness (MINECO). 
FSL was granted access to the HPC resources of The Institute for
scientific Computing and Simulation financed by Region Ile de France
and the project Equip@Meso (reference ANR-10-EQPX- 29-01) overseen by
the French National Research Agency (ANR) as part of the
Investissements d'Avenir program. 
This work has made use of data from the European Space Agency (ESA) mission
{\it Gaia} (\url{https://www.cosmos.esa.int/gaia}), processed by the {\it Gaia}
Data Processing and Analysis Consortium (DPAC,
\url{https://www.cosmos.esa.int/web/gaia/dpac/consortium}). Funding for the DPAC
has been provided by national institutions, in particular the institutions
participating in the {\it Gaia} Multilateral Agreement.

\end{acknowledgements}

\appendix

\section{Lucy's method for the inversion of Fredholm integral equations 
of the first kind}
\label{.Lucy}

The inversion of Fredholm integral equations of the first kind such as Eq. (\ref{convol}) 
is ill-conditioned. Typical analytical methods for solving these equations (see Bal\'azs 1995) cannot achieve a good solution because of the sensitivity of the kernel to the noise of the star counts (see, e.g., Craig \& Brown 1986, chapter 5).
Since the functions in these equations have a stochastic rather than analytical interpretation, it is to be expected that statistical-inversion algorithms will be more robust (Turchin et al. 1971; Jupp \& Vozoff 1975; Balázs 1995).
Among these statistical methods, the iterative method of Lucy's algorithm 
(Lucy 1974; Turchin et al. 1971; Bal\'azs 1995; L\'opez-Corredoira et al. 2000) is an
appropriate one. Its key feature is the interpretation of the kernel as a conditioned probability and the application of Bayes's theorem.

In Eq. (\ref{convol}), $N(\pi )$ is the unknown function, and the kernel is $G(x)$, whose
difference $x$ is conditioned to the parallax $\pi '$. The inversion is carried out as:
\begin{equation}
N (\pi )=\lim _{n\rightarrow \infty}N _{n}(\pi )
,\end{equation}
\begin{equation}
N_{n+1}(\pi)=N_n(\pi )\frac{\int _0^\infty \frac{\overline{N}(\pi ')}
{\overline{N_n}(\pi ')}G _{\pi '}(\pi -\pi')d\pi '}
{\int _0^\infty G_{\pi '}(\pi -\pi')d\pi '}
,\end{equation}
\begin{equation}
\overline{N_n}(\pi )=\int _0^\infty N_n(\pi ')G_{\pi '}(\pi -\pi')d\pi '
.\end{equation}
The iteration converges when $=\overline{N_n}(\pi )\approx \overline{N}(\pi )$ 
$\forall \pi$, that is, when $N _{n}(\pi )\approx N (\pi )$ $\forall \pi$. 
The first iterations produce a result that is close to the final answer, with the subsequent iterations giving only small corrections. In our calculation, we set as
initial function of the iteration $N_0(\pi )=\overline{N}(\pi )$ and we carry out
a number $n\ge 3$ of iterations until the Pearson's $\chi ^2$ test gives
\begin{equation}
\frac{1}{N_p-2}\sum _{j=2}^{N_p-1}\frac{[\overline{N_n}(\pi _j)-\overline{N}(\pi _j)]^2}{\overline{N_n}(\pi _j)}<1
,\end{equation}
that is, evaluating the convergence within the Poissonian noise
in the $N_p$ examined values of $\pi $ except the border points.
Further iterations would enter within the noise.

This algorithm has a number of beneficial properties (Lucy 1974, 1994): all the functions are defined as being positive, the likelihood increases with the number of iterations, the method is insensitive to high-frequency noise in $\overline{N}(\pi )$, and so on.
We note however that, precisely because this method only works when $N$ are positive functions, it does not work with negative ones.

\end{document}